**Annama H chondrite – mineralogy, physical properties, cosmic ray exposure, and parent body history**

Tomáš Kohout[1,2,3], Jakub Haloda[4], Patricie Halodová[4], Matthias M. M. Meier[5], Colin Maden[5], Henner Busemann[5], Matthias Laubenstein[6], Marc. W. Caffee[7,8], Kees C. Welten[9], Jens Hopp[10,11], Mario Trieloff[10,11], Ramakant R. Mahajan[12], Sekhar Naik[12], Josep M. Trigo-Rodriguez[13], Carles E. Moyano-Cambero[13], Michael I. Oshtrakh[14], Alevtina A. Maksimova[14], Andrey V. Chukin[14], Vladimir A. Semionkin[14], Maksim S. Karabanalov[15], Israel Felner[16], Evgeniya V. Petrova[14], Evgeniia V. Brusnitsyna[14], Victor I. Grokhovsky[14], Grigoriy A. Yakovlev[14], Maria Gritsevich[1,3,14,18], Esko Lyytinen[3], Jarmo Moilanen[3], Nikolai A. Kruglikov[14,17], Aleksey V. Ishchenko[14]

1 Department of Physics, University of Helsinki, Finland. E-mail: tomas.kohout@helsinki.fi.

2 Institute of Geology, The Czech Academy of Sciences, Prague, Czech Republic.

3 Finnish Fireball Network, Finland.

4 Czech Geological Survey, Prague, Czech Republic

5 Institute of Geochemistry and Petrology, ETH Zurich, Switzerland.

6 Laboratori Nazionali del Gran Sasso, Istituto Nazionale di Fisica Nucleare, Via G. Acitelli 22, I-67100 Assergi (AQ), Italy

7 Department of Physics and Astronomy, Purdue University, West Lafayette IN, 47907, USA

8 Department of Earth, Atmospheric, and Planetary Sciences, Purdue University, West Lafayette, IN 47907, USA

9 Space Sciences Laboratory, University of California, Berkeley, CA 94720, USA

10 Institut für Geowissenschaften, Universität Heidelberg, Im Neuenheimer Feld 234-236, D-69120 Heidelberg, Germany.

11 Klaus-Tschira-Labor für Kosmochemie, Im Neuenheimer Feld 234-236, D-69120 Heidelberg, Germany.

12 Physical Research Laboratory, Ahmedabad – 380009, India

13 Institute of Space Sciences (IEEC-CSIC), Meteorites, Minor Bodies and Planetary Sciences Group Campus UAB Bellaterra, c/Can Magrans s/n, 08193 Cerdanyola del Vallès (Barcelona), Spain.

14 Institute of Physics and Technology, Ural Federal University, Ekaterinburg, 620002, Russia.

15 Institute of Material Science and Metallurgy, Ural Federal University, Ekaterinburg, 620002, Russia.

16 Racah Institute of Physics, The Hebrew University, Jerusalem, Israel

17 Institute of Metal Physics, Ural Branch of Russian Academy of Sciences, Russia.

18 Dorodnicyn Computing Centre, Federal Research Center Computer Science and Control of the Russian Academy of Sciences, 119333 Moscow, Russia.

**Abstract**

The fall of the Annama meteorite occurred early morning (local time) on April 19, 2014 on the Kola Peninsula (Russia). Based on mineralogy and physical properties, Annama is a typical H chondrite. It has a high Ar-Ar age of 4.4 Ga. Its cosmic ray exposure history is atypical as it is not part of the large group of H chondrites with a prominent 7 – 8 Ma peak in the exposure age histograms. Instead, its exposure age is within uncertainty of a smaller peak at 30 ±4 Ma. The results from short-lived radionuclides are compatible with an atmosperic pre-entry radius of 30 – 40 cm. However, based on noble gas and cosmogenic radionuclide data, Annama must have been part of a larger body (radius >65 cm) for a large part of its cosmic ray exposure history. The $^{10}$Be concentration indicates a recent (3 – 5 Ma) breakup which may be responsible for the Annama parent body size reduction to 30 – 35 cm pre-entry radius.

**Introduction**

The fall of the Annama meteorite occurred early morning (local time) on April 19, 2014 on the Kola Peninsula (Russia), close to the Finnish border. The event is also known as the Kola fireball (*Gritsevich et al. 2014a)*. The fireball was observed by the Finnish Fireball Network as well as by numerous eyewitnesses in Finland and local residents. The strewnfield was computed from the observations by the Finnish Fireball Network and a calibrated video (taken by Alexandr Nesterov in Snezhnogorsk, Russia), details are provided in *Gritsevich et al. (2014b)*. The fireball observations and heliocentric orbit reconstruction are detailed in *Trigo-Rodríguez et al. (2015), Dmitriev et al. (2015)*. The fireball data interpretation was done using the fb_entry program (*Lyytinen and Gritsevich 2013*) taking into account the atmospheric conditions present during the meteorite fall (*Lyytinen and Gritsevich 2016*). The derived characteristic parameters as well as the terminal height of the fireball indicated the possibility of surviving meteorite fragments reaching the ground (*Gritsevich et al. 2012, Moreno-Ibáñez et al. 2015*). The pre-entry mass of the Annama meteoroid was estimated based on the deceleration analysis as described by *Gritsevich (2008, 2009)* to be in the range of 400 – 500 kg (*Trigo-Rodríguez et al. 2015*, *Lyytinen and Gritsevich 2016*), with corresponding radius around 30 – 35 cm.

Annama collided with the Earth on an Apollo-type orbit (*Trigo-Rodríguez et al. 2015, Dmitriev et al. 2015*) and at the time of its fall it was the 23rd meteorite and 10th H chondrite with known heliocentric orbit (http://www.meteoriteorbits.info/, retrieved November 10, 2016).

In May 2014, a meteorite recovery expedition was organized to the predicted landing area (*Gritsevich et al. 2014b*). Two meteorites (Fig. 1) referred to as Annama I (120.4 g) and Annama II (47.5 g) were found only about 100 meters from the predicted fall line on May 29 and May 30, respectively.

**Materials and methods**

*Optical and scanning electron microscopy*

Two polished thin sections (265 and 220 mm$^2$) of Annama I were prepared for study of the meteorite petrography, mineralogy and micro-chemical composition. For final polishing Struers OPU colloidal silica suspension was used. The first thin section (265 mm$^2$) was coated by 25 nm thick conductive layer of carbon for the purposes of microanalytical studies. The second thin section was used also to study optical features by a LEICA DLMP petrographic microscope.

A scanning electron microscope (SEM) TESCAN MIRA 3GMU at the Czech Geological Survey, Czech Republic, was used for microanalytical studies and collection of backscattered electron images. Mineral compositions were determined using an Oxford Instruments AZtec Energy Automated analytical system with a SDD X-ray energy dispersive detector X-Max$^N$ 80 Premium (EDS) and a X-ray wavelength dispersive detector Wave 700 (WDS). Both analytical systems installed on the SEM were used for simultaneous EDS-WDS analysis. Analyses were performed using an accelerating voltage of 15 kV, 20 nA beam current, 0.120 µm beam size and XPP matrix correction procedures. Acquisition live time for analysis of major elements by EDS was 60 s and the output count rate was 1.1 x 10$^5$ counts per second. Counting times for analysis of minor and trace elements by WDS system were 20 or 30 s. To avoid Na and K ions migration during the analysis of plagioclase, the beam diameter was enlarged to 5 µm. For simultaneous EDS-WDS analysis a combination of natural and synthetic standards was used for standardization and calibration process.

An integrated electron backscattered diffraction system (EBSD) Oxford Instruments AZtec HKL Automated and Nordlys Nano detector, installed on abovementioned SEM was used for confirmation of microstructural characteristics of the studied minerals. Analytical conditions were 15 mm working distance, 20 kV accelerating voltage and 3.5 nA beam current.

For metallographic purposes, a 220 mm$^2$ polished section of Annama I was studied at the Ural Federal University, Russia. The structure of metal particles was revealed by etching with 2% $HNO_3$ solution. Samples were studied using an Axiovert 40 MAT (optical microscopy) and a Carl Zeiss ΣIGMA VP (SEM). Optical data were processed using a Panoramic Microscopy System SIAMS MT.

*X-ray diffraction*

X-ray diffraction (XRD) patterns were measured on a 88 mg powdered sample of Annama II at the Ural Federal University using a XRD-700 diffractometer (Shimadzu) operated at 40 kV and 30 mA with $CuK_\alpha$ radiation and silicon monochromator, scanned over 2Θ from 13° to 85° with a step of 0.05° per 7 s.

*Raman spectroscopy*

Micro-Raman spectra were recorded at the Institut de Ciència de Materials de Barcelona (ICMAB - CSIC) in backscattering geometry at room temperature using the 5145 Å line of an Argon-ion laser with a Jobin-Yvon T-64000 Raman spectrometer attached to an Olympus microscope and equipped with a liquid nitrogen-cooled CCD detector. 71 spots of different regions and minerals were analyzed with this technique. Analysis spot was ~1 µm and the laser power on the sample was kept below 0.5 mW to avoid degradation. This setup allowed for high-resolution spectra acquisition in range of 100 – 1400 cm$^{-1}$.

*Density and porosity*

The bulk volume was determined at the Department of Physics, University of Helsinki, using a modified Archimedean method (*Consolmagno and Britt 1998*, *Macke et al. 2010*) incorporating glass beads ~ 0.3 mm in diameter. Ten sets of measurements per sample were performed. The method was thoroughly tested and calibrated prior to the measurements using volume standards and the absolute resolution and precision were determined to be ±0.1 $cm^3$. The grain volume was measured using a Quantachrome Ultrapyc 1200e He pycnometer. The absolute resolution and the precision of this device are estimated to be ±0.05 $cm^3$. Mass was determined using a digital OHAUS Navigator scale with 0.1 g resolution and precision. The scale was calibrated prior to the measurements using internal calibration.

*Magnetic susceptibility*

The room temperature magnetic susceptibility was measured in the field upon recovery of the meteorites using a ZH instruments SM-30 portable surface susceptibility meter with a large 5 cm coil operating at 8 kHz frequency. The values were corrected for the actual sample size following the procedure described in *Gattacceca et al. (2004)*. Later, the susceptibility was verified in the laboratories at the University of Helsinki and at the Ural Federal University using a ZH instruments SM-100 susceptibility meter operating at 8 kHz frequency and 320 A/m RMS field amplitude.

The susceptibility of the samples was measured three times along three perpendicular directions. Subsequently, a logarithm of the average apparent magnetic susceptibility (in $10^{-9}$ $Am^2/kg$) was calculated as described in *Rochette et al. (2003)*. The relative error in the determined value of the magnetic susceptibility logarithm is below 3%.

*Gamma-ray spectroscopy (GRS)*

The concentrations of short- and medium-lived cosmogenic radionuclides, as well as long-lived cosmogenic $^{26}$Al and natural radioactivity ($^{40}$K, $^{232}$Th, $^{235}$U), were measured on a 6.58 g sample from Annama II using non-destructive gamma-ray spectroscopy in the STELLA (SubTErranean LowLevel Assay) facility of underground laboratories at the Laboratori Nazionali del Gran Sasso (LNGS) in Italy, using a high-purity germanium (HPGe) detector of 518 $cm^3$ (*Arpesella 1996*). The measurement was done 101 days after the fall and the counting time was 34.61 days. The counting efficiencies were calculated using a Monte Carlo code. This code is validated through measurements and analyses of samples of well-known radionuclide activities and geometries. The uncertainties in the radionuclide activities are dominated by the uncertainty in the counting efficiency, which is conservatively estimated at 10%. Because actual data for Annama were not yet available, the average H Chondrite density and composition were taken from *Britt and Consolmagno (2003)*, and from *Jarosewich (1990)*, respectively. Exact values determined later for Annama are almost identical and thus this approach has no influence on results.

*Accelerator mass spectrometry (AMS)*

A 68 mg sample from Annama I was used for cosmogenic radionuclide separation at the UC Berkeley's Space Sciences Lab and subsequent measurement at PRIME Lab at Purdue University. For the analysis of cosmogenic $^{10}$Be (half-life 1.36 x $10^6$ a), $^{26}$Al (7.05 x $10^5$ a) and $^{36}$Cl (3.01 x $10^5$ a), we crushed the 68 mg aliquot in an agate mortar and separated the magnetic (metal) fraction (~15 mg) and non-magnetic (stone) fraction (~50 mg) with a magnet. We dissolved 49.9 mg of the stone fraction, along with 2.95 mg of Be and 3.70 mg of Cl carrier, in concentrated $HF/HNO_3$. After dissolution, we separated the Cl fraction as AgCl and purified the AgCl for isotopic analysis by AMS. The solution was evaporated to dryness, thereby removing Si as $SiF_4$. After fuming with $HClO_4$, the residue was dissolved in dilute HCL, and a ~3% aliquot was taken for chemical analysis by ICP-OES. We added 5.35 mg of Al carrier and separated Be and Al using ion exchange and acetyl-acetone extraction methods described previously (*Welten et al., 2011*). The $^{10}Be/Be$, $^{26}Al/Al$ and $^{36}Cl/Cl$ ratios of the samples were measured by accelerator mass spectrometry at the Purdue University PRIME Lab (Sharma et al. 2000). The measured ratios were corrected for blanks (<1% correction) and normalized to $^{10}Be$, $^{26}Al$ and $^{36}Cl$ AMS standards (*Sharma et al.,* 1990, Nishiizumi *2004*, *Nishiizumi et al. 2007*).

*Noble gases and nitrogen*

The first round of noble gas (He, Ne, Ar) analyses was done at the noble gas laboratory of ETH Zurich, Switzerland, on samples from Annama I, referred in the text as Annama Ia (114 mg) and Annama Ib (54 mg) using a protocol most recently described in detail by *Meier et al. (2017)*. Samples were pre-heated at 100°C for 24 hours in ultra-high vacuum ($10^{-10}$ mbar) to remove potentially adsorbed atmospheric gases. The gas was then extracted in a single temperature step at ~1700°C in a furnace heated by electron bombardment. He and Ne were separated from Ar using a cold trap held at the temperature of liquid nitrogen. Sample measurements were bracketed by blank measurements, and blanks subtracted from the sample and calibration gas signals. The blank contribution to the released sample gas was <1% for all He, Ne isotopes, and <15% for Ar isotopes.

The second round of noble gas (He, Ne, Ar, Kr and Xe) analyses as well as a nitrogen analysis was done on a 74.69 mg sample of Annama I referred in the text as Annama Ir. The mass spectrometric measurements were performed on a 'Noblesse' (Nu Instruments U.K.) multi collector noble gas mass spectrometer at Physical Research Laboratory, Ahmedabad, India using standard procedures (*Mahajan 2015*, *Mahajan et al. 2016*). Extraction was performed in four temperature steps, 600°C, 900°C, 1200°C and 1700°C in a resistance furnace. The blank levels (at 1700°C, in units of $10^{-11}$ $cm^3$STP) were 223, 4.89, 3.39, 0.0457, and 0.000166 for $^{4}He$, $^{22}Ne$, $^{36}Ar$, $^{84}Kr$ and $^{132}Xe$, respectively. For nitrogen, the blank contribution to the total signal was ~2%.

*Ar-Ar dating*

A 48.8 mg whole rock sample from Annama I was cleaned with pure ethanol and wrapped into high-purity Al-foil and placed in an evacuated quartz vial, bracketed by NL25 hornblende age standards (2657 ±4 Ma, *Schwarz and Trieloff 2007*). In addition, one $CaF_2$-standard was added to account for Ca interference reactions. Subsequently, the sample was neutron-irradiated with Cd-shielding at the Portuguese Research Reactor in Bobadela for 96 h (corresponds to a neutron flux of ca. 2·x $10^{18}$ n/cm²). Argon analyses followed the procedure given by *Hopp et al. (2014)*. Blank for the $^{40}Ar$ were interpolated for each temperature step from measured blank assuming an exponential increase in gas amount. Observed blank concentrations of $^{40}Ar$ were (in $10^{-10}$ cm³ STP) 1.25 ±0.02 at 800°C, 6.30 ±0.04 at 1200°C, 34.5 ±0.2 at 1400°C and 325.5 ±1.7 at 1700°C. The isotopic composition of the blank was indistinguishable from atmospheric composition within uncertainty. For the 700°C extraction we applied a significantly elevated blank correction of 90±9·$10^{-10}$ cm³ STP to account for an additional blank during melting of the Al-foil. This value is an interpolation and assumes a linear change in observed $^{38}Ar/^{36}Ar$ and $^{40}Ar/^{36}Ar$-ratios of successive extractions. For evaluation we used the recommended decay parameters given in *Steiger and Jäger (1977)*. Note, that at 4.5 Ga the age bias between the U-Pb and K-Ar timescales may cause an correction towards older ages of about 20 – 30 Ma of our reported Ar-Ar age (*Renne et al. 2011*, *Schwarz et al. 2011*). Correction parameters for interference reactions with K and Ca were $(^{38}Ar/^{39}Ar)_K$ = 0.01465 ±0.00100, $(^{40}Ar/^{39}Ar)_K$ = 0.0123 ±0.0024, $(^{36}Ar/^{37}Ar)_{Ca}$ = 0.00030 ±0.00016, $(^{39}Ar/^{37}Ar)_{Ca}$ = 0.00079 ±0.00001, $(^{40}Ar/^{37}Ar)_{Ca}$ = 0.003 ±0.003. The J-value was 0.004741 ±0.000030 (internal error) / ±0.000033 (external error). Results of our Ar-Ar analyses are listed in Table S1 in Supporting Information. The data are blank and interference corrected (for K and Ca) but still not deconvolved into spallation and trapped components. However, for isochron calculations correction for spallation was applied.

*Mössbauer spectroscopy and magnetization measurements*

An 88 mg powdered sample of Annama II glued on an iron free Al-foil (the sample had a thickness of ~8 mgFe/cm$^2$) was used for Mössbauer spectroscopy with high velocity resolution. The Mössbauer spectrum was measured at the Ural Federal University using an automated precision Mössbauer spectrometric system built on the base of the SM-2201 spectrometer with a saw-tooth shape velocity reference signal formed by the digital-analog converter with discretization of $2^{12}$ (quantification with 4096 steps). Details of the instrument can be found in *Oshtrakh et al. (2009)*, *Semionkin et al. (2010)*, and *Oshtrakh and Semionkin (2013)*. The ~ $10^9$ Bq $^{57}$Co(Rh) source (Ritverc GmbH, St. Petersburg) was used at room temperature. The measurement was carried out in transmission geometry with a moving absorber at 295 K and recorded in 4096 channels. Then this spectrum was converted into the 1024-channel spectrum by a consequent summation of four neighboring channels to increase the signal-to-noise ratio for the minor spectral components. The statistical count rate in the 1024-channel Mössbauer spectrum of Annama H5 ordinary chondrite was 19.6 x $10^6$ counts per channel and the signal-to-noise ratio was 239. The Mössbauer spectrum was fitted with the least squares procedure using the UNIVEM-MS program with a Lorentzian line shape. A new approach to fit the troilite magnetic spectral component using the simulation of the full static Hamiltonian was applied as already reported by *Maksimova et al. (2016b, 2017*). Spectral parameters such as isomer shift (δ), quadrupole splitting (quadrupole shift for magnetically split components) ($\Delta E_Q$), magnetic hyperfine field ($H_{eff}$), line width (Γ), relative area of spectral components (A), and statistical criterion ($\chi^2$) were determined. An instrumental (systematic) error for each spectrum point was ±0.5 channel (in mm/s), the instrumental (systematic) error for the hyperfine parameters was ±1 channel (in mm/s or kOe). If the statistical error calculated with the fitting procedure (fitting error) for these parameters exceeded the instrumental (systematic) error the larger error was used instead. The criteria for the best fit were differential spectrum, $\chi^2$, and physical meaning of the spectral parameters. Values of δ are given relative to α-Fe at 295 K.

The Mössbauer data analysis was further supported by optical and electron microscope observations of a small thin section from Annama II carried out by means of an Axiovert 40 MAT optical microscope (Carl Zeiss) and an Auriga CrossBeam SEM (Carl Zeiss) with an X-max 80 EDS device (Oxford Instruments) at the Ural Federal University.

Field cooled (FC) and zero field cooled (ZFC) curves of induced magnetization (at 250 Oe field) as well as isothermal magnetization M(H) curves (at 5 K) were measured using a SQUID magnetometer MPMS-5S (Quantum Design) at the Hebrew University, Israel on a 6.2 mg powdered sample from Annama II.

**Results**

*Petrography, mineralogy, shock features, and physical properties*

Upon recovery, the 120.4 g Annama I meteorite was ~70% covered with black fusion crust and with apparent stream lines on one side (Fig. 1). The fresh surface was bright with abundant thin dark impact melt veins. The 47.5 g Annama II meteorite was fully covered with dark fusion crust.

Annama has a recrystallized fine-grained granular texture. Chondrules (200 – 3000 µm in diameter) are composed predominantly of olivine and pyroxene and their shapes are mostly readily delineated (Fig. 2). Their types include radial pyroxene (RP), barred olivine (BO), and cryptocrystalline (CC) chondrules. Irregular fractures in olivine and undulatory extinction of olivine and plagioclase (no maskelynite was detected by EBSD) indicate a shock stage S2 based on the scale by *Stöffler et al. (1991)*. Thin veins of impact melt are also present. No weathering products were observed (W0).

EDS-WDS show an olivine composition Fa18.6 ±0.3 (N=60) and low-Ca pyroxene composition Fs16.6 ±0.2 and Wo1.26 ±0.26 (N=65). Also present are diopside (Fs6Wo46), plagioclase (Ab80An14Or6), troilite, kamacite, taenite, tetrataenite, and chromite. Accessory minerals include chlorapatite, hercynite, ilmenite, merrillite, and native copper. Classification of the chondrite as type H5 was made following criteria published in *Huss et al. (2006)* (Fig. 3).

The XRD pattern of the Annama H5 fragment is shown in Fig. 4. An analysis of XRD data using full profile Rietveld analysis shows the presence of the following phases: olivine (~38.6 wt%), orthopyroxene (~36.6 wt%), α-Fe(Ni, Co) phase (~9.0 wt%), troilite (~5.6 wt%), anorthite (~4.7 wt%), chromite (~2.7 wt%), clinopyroxene (~1.4 wt%), γ-Fe(Ni, Co) phase (~1.3 wt%) and hercynite (~0.2 wt%). The unit cell parameters for olivine are: a=10.2476(9) Å, b=6.0026(7) Å, c=4.7671(5) Å and for orthopyroxene are: a=18.2746(8) Å, b=8.8493(7) Å, c=5.2044(7) Å. The $Fe^{2+}$ occupancies of the M1 and M2 sites in silicates were also determined from the XRD pattern analysis. These values are $X_{Fe}^{M1}$=21 and $X_{Fe}^{M2}$=17 for olivine and $X_{Fe}^{M1}$=6 and $X_{Fe}^{M2}$=48 for orthopyroxene.

Metallographic image analysis of 2740 metal and troilite grains shows that relative metal area is 6.01% (corresponding to approx. 15 – 20 wt%) and relative troilite area is 1.66%. Troilite is mostly present as small grains as one can see from distribution of metal and troilite particle sizes shown in Fig. 5. The metal grains are composed predominantly (96.5%) of kamacite, while zoned (cloudy taenite) metal is sparse (3.5%). Kamacite forms larger grains (largest observed 5 x 1.5 mm), while zoned taenite particles are usually smaller, have complex structure and associate with troilite.

Pure metallic copper inclusions were observed as individual grains in contact with troilite ( Fig. 6) or in a form of precipitations inside zoned taenite particles. This structure is typical for the zoned polycrystalline particles within H chondrites (as observed for example in Okhansk H4, *Grokhovsky 1988*).

All zoned taenite particles contain cloudy zone formed during spinodal decomposition in metal areas with 40 – 45% of Ni. The cooling rate of this metal grains could be defined by the Ni-rich phase average size within the cloudy taenite zone as described by *Goldstein (2009)*. The average size of these particles near the tetrataenite phase is 80 nm (Fig. S1 in Supporting Information), which corresponds to a cooling rate of about 20 K/Ma below 350°C. The same values were obtained earlier for the ordinary chondrites Forest City (H5, S2), Landreth Draw (H5, S2), Richardton (H5, S1), Sena (H4, S1) by *Scott et al. (2014)*.

Raman spectroscopy confirms the results from EMPA. From the 71 analyzed spots, most correspond, as expected, to olivine (peaks at ~822 and 853 $cm^{-1}$). Pyroxene (peaks at ~333, 393, 654, 674, and 1002 $cm^{-1}$), plagioclase (peaks at ~473 and 503 $cm^{-1}$), troilite (peaks at ~213, 279 and 391 $cm^{-1}$) and chromite (peak at ~690 $cm^{-1}$) are also detected as relatively common phases. From the accessory minerals we focused on merrillite (peaks at ~958 and 973 $cm^{-1}$). Strong enough shock events would have caused the transformation of merrillite in veins into polymorph with structure of trigonal $\gamma$-$Ca_3(PO_4)_2$ (*Xie et al. 2002*). This phase was not detected, indicating that the sample is not strongly shock-metamorphosed, which is consistent with the low S2 shock degree defined for Annama.

The bulk and grain density of both meteorites are 3.5 $g/cm^3$ and 3.8 $g/cm^3$ respectively. Resulting porosity is 8%. Logarithm of magnetic susceptibility (measured in $10^{-9}$ $m^3/kg$) of both meteorites is 5.4. The density and magnetic susceptibility values are also consistent with H chondrite.

The complete description of the Annama fall, recovery, physical appearance, and classification is available from Meteorite Bulletin, 104, 2015 (https://www.lpi.usra.edu/meteor/metbull.php?code=61696).

*Cosmogenic radionuclides*

We report here long-lived cosmogenic radionuclides in Annama I (Table 1) and short-lived radionuclides plus $^{26}Al$ in Annama II (Table 2). Short-, medium-, and long-lived radionuclides detected in Annama II are $^{26}Al$, $^{7}Be$, $^{22}Na$, $^{56}Co$, $^{57}Co$, $^{58}Co$, $^{60}Co$, $^{54}Mn$, $^{46}Sc$. For $^{44}Ti$ only upper detection limits are reported. The given activities are calculated for the date of fall following the simple decay law, and taking into account the time that passed between the fall of the meteorite and its measurement.

*Cosmogenic, radiogenic, and trapped noble gases, and nitrogen*

The noble gas (He, Ne, Ar) concentrations and isotopic ratios measured in samples of Annama Ia and Ib (used for the determination of the cosmic history) are listed in Table 3. The step-wise heating data for He, Ne, Ar, and Kr from Annama Ir are listed in Table S2 in Supporting Information, while the step-wise heating data for Xe are given in Table S3 in Supporting Information. In all three Annama I samples, He, Ne and Ar are dominated by cosmogenic and radiogenic components (from cosmic ray induced spallation reactions, and radioactive decay, respectively). Helium is released at lower temperatures (mostly ≤900°C) compared to Ne, Ar, Kr and Xe (mostly >900°C; Tables S1 and S2 in Supporting Information). The low cosmogenic $^{22}Ne/^{21}Ne$ ratio of 1.058 ±0.003 suggests that Annama acquired most of its cosmogenic noble gases in a body which was at least 65 cm in radius, significantly larger than the one delivering the Annama meteorite in 2014, hinting at a complex multi-stage cosmic ray exposure age. Combining the cosmogenic $^{3}He$, $^{21}Ne$ and $^{38}Ar$ concentrations with the shielding information from the $^{22}Ne/^{21}Ne$, cosmogenic radionuclides and a production rate model (*Leya and Masarik 2009*, *Dalcher et al. 2013*), we obtain cosmic ray exposure ages of 28, 34, and 27 Ma for He, Ne, and Ar, respectively (samples Annama Ia + Ib). The radiogenic $^{4}He$ and $^{40}Ar$ can be combined with the typical abundances of K, U, Th in H chondrites (*Wasson and Kallemeyn 1988*) to determine a nominal U, Th-He age of 2.7 Ga for Annama Ia + Ib, and 2.5 Ga for Annama Ir, compatible within the typical uncertainty of this method. The K-Ar age of all three Annama I samples is identical within uncertainties at 3.8 Ga.

No solar-wind-derived He and Ne was found, therefore Annama is not a regolith breccia (about 15% of all H chondrites are regolith breccias, *Bischoff and Schultz 2004*), nor are there other trapped components present for He and Ne. For Ar, a small contribution from a trapped (t) component (likely Ar-Q given the Q-like $^{36}Ar_t/^{132}Xe_t$ and $^{84}Kr_t/^{132}Xe_t$ ratios, Fig. S2 in Supporting Information) is present. Kr and Xe are dominated by the Q component (*Busemann et al. 2000*), with a concentration in the typical range of an equilibrated ordinary chondrite. Besides Q gases, there is also a contribution of cosmogenic Kr ( Fig. S3 in Supporting Information). A small excess of $^{82}$Kr by neutron-induced reactions on Br are visible in the $^{83}Kr/^{84}Kr$ vs $^{82}Kr/^{84}Kr$ three isotope plot (Fig. S4 in Supporting Information). The neutron-induced Kr excess is very small, either due to very little shielding, or very large shielding deep inside a large parent body (>2 m). The second interpretation is in agreement with the high shielding conditions derived from the low $^{22}Ne/^{21}Ne$ ratio. Some $^{129}$Xe from the decay of now-extinct $^{129}$I ($t_{1/2}$ = 15 Ma) is also present, with a $^{129}Xe/^{132}Xe$ ratio of 1.144.

The nitrogen isotopic composition is reported as $\delta^{15}N$ (‰) where: $\delta^{15}N=[\{(^{15}N/^{14}N)_{sample}/(^{15}N/^{14}N)_{standard}\}-1]\times1000$, and the standard is atmospheric nitrogen ($^{15}N/^{14}N = 0.003676$). In Annama Ir, 39.2 ppm nitrogen was released (most of it at the 900°C step, see Table S2) and the measured isotopic signature have $\delta^{15}N = 4.37 \pm 0.13$‰. These values are in the typical range of ordinary chondrites (*Hashizume and Sugiura 1995*, *Sugiura et al. 1998*). Cosmogenic $^{15}$N is produced in meteorites through spallation reactions (*Mathew and Murty 1993*). The trapped (t) nitrogen after cosmogenic (c) correction, calculated as $\delta^{15}N_t = (\delta^{15}N_m-\delta^{15}N_c)$ using cosmogenic $^{21}Ne_c$ (*Mathew and Murty 1993*) has $\delta^{15}N_t = 0.97$ ‰.

*Ar-Ar dating*

The age spectrum of Annama (Fig. 7) shows no definite age plateau, if evaluated with a trapped solar $^{40}Ar/^{36}Ar$ ratio of 1 ±1 (i.e. essentially all $^{40}$Ar is radiogenic). At higher degassing temperatures we have evidence of recoil-redistribution of $^{39}$Ar from K-rich lower temperature phases into K-poor higher temperature phases, resulting in lowered apparent ages. This also agrees with an elevated age at 560°C which cannot be readily explained by a different and still unaccounted trapped Ar component and more likely mirrors the preferential loss of $^{39}$Ar via recoil from finer-grained K carriers. The respective integrated age of Annama is 4399 ±11 (int.) / 12 (ext.) Ma. A closer inspection of the data in a $^{40}Ar/^{36}Ar$ versus $^{39}Ar/^{36}Ar$ diagram (Fig. 8) revealed a linear relation at 675 – 780°C which can be interpreted as an isochron. The intercept of this isochron is 404 ±45 Ma, and the $^{40}Ar/^{39}Ar$ ratio derived by the slope would correspond to an age of 4426 ±13 (int.) / 14 (ext.) Ma which agrees well with an inverse isochron age of 4418 ±11 (int.) / 12 (ext.) Ma and with a plateau age calculation of 4420 ±12 (int.) / 13 (ext.) Ma ( Fig. 7). Isochron regression was performed with the Isoplot 4.15 program (*Ludwig 2008*) with correlated errors (equation 14 in *Ludwig and Titterington (1994)*). Note, that we used the uncertainties given for the observed scatter.

Major release of Ca-derived $^{37}$Ar was detected at higher temperatures (1080 – 1400°C), but no age plateau defining the $^{38}$Ar cosmic ray exposure age could be identified. Assuming a production ratio of $2.8\cdot10^{-8}$ $cm^3$ STP $^{38}$Ar /[g Ca Ma] (*Leya and Masarik (2009)*) equivalent to the production rate at the center of a meteoroid with a radius of 65 cm the lowest calculated cosmic ray exposure ages are 39.8 ±0.5 Ma at 1220°C and 41.7 ±2.9 Ma at 1300°C. Note, that both extractions encompass 45% of the total $^{37}$Ar release. The ages are slightly above the estimates based on $^{21}$Ne or $^{38}$Ar production rates (34 Ma and 27 Ma, respectively) derived by noble gas analyses (see subsection "Cosmic ray exposure and thermal history" in Discussion section) but must be considered as an upper limit since production of spallogenic $^{38}Ar$ from Fe and Ni is not included.

*Mössbauer spectroscopy and magnetization measurements*

The Mössbauer spectrum of Annama is shown in Fig.9. This spectrum visually consists of several magnetic sextets and quadrupole doublets. The best fit of the spectrum revealed the presence of 8 magnetic sextets, 5 quadrupole doublets and 2 singlets (parameters are given in Table S4 in Supporting Information).

Magnetic sextets 1–7 are related to different Fe-Ni-Co alloy phases whereas sextet 8 is related to troilite (FeS). Judging from their hyperfine parameters sextet 1 can be related to $\alpha_2$-Fe(Ni, Co) phase while sextets 2–5 may be associated with the same $\alpha$-Fe(Ni, Co) phase but with variations in Ni and Co concentration. Sextets 6 and 7 can be assigned to $\gamma$-Fe(Ni, Co) and $\gamma$-FeNi, respectively. Quadrupole doublets 9 and 10 and 11 and 12 are related to the M1 and M2 sites in olivine and orthopyroxene, respectively. Additional minor quadrupole doublet 13 is associated with hercynite. Two singlets 14 and 15 are related to paramagnetic $\gamma$-Fe(Ni, Co) and chromite phases, respectively. The abovementioned phases were confirmed by SEM-EDS observations at the Ural Federal University. The values of hyperfine parameters of these spectral components (Table S4) are in agreement with those recently obtained for corresponding phases in H, L and LL ordinary chondrites *(Maksimova et al. 2016ac, Oshtrakh et al. 2016, Maksimova et al. 2017*).

The FC and ZFC curves and M(H) curve measured for Annama H5 meteorite fragment are shown in Fig. 10. These curves clearly indicate a multi-phase bulk material which may be composed of paramagnetic (PM) and magnetically ordered components. The low temperature magnetization decrease in both curves reflects the PM phases whereas the increase at the high temperature range is attributed to the various magnetic phases. Thus, the low temperature inflections are the compensation points of these two components. The 60 K feature on the FC/ZFC curve can be associated with the Curie temperature of chromite (*Gattacceca et al. 2011*). The nature of the bulges at around 70–80 K and 200 K are not clear. FC/ZFC curves did not converge at the temperature range up to 300 K indicating a higher blocking temperature of the remanence carrier.

The isothermal magnetization M(H) curve measured at 5 K first increases linearly up to 2 kOe and then tend to saturate. M(H) is also a mixture of two components and can be fitted as: $M(H) = M_s+\chi_p H$, where $M_s$ = 38.9 emu/g is the saturation magnetization of the intrinsic magnetic phases and $\chi_p H$ is the linear PM part. The part of the hysteresis curve at 5 K (Fig. 10) shows a remanence $M_{rs}$ value of 0.48 emu/g and a small coercive field $H_c$ of 16 Oe which are the bulk values of the various Fe(Ni, Co) magnetic phases.

## Discussion

*Mineralogy and physical properties*

Mineralogical and physical properties data indicate that Annama is a rather typical H chondrite with moderate shock and low weathering degree. The measured bulk (3.5 g/cm$^3$) and grain density (3.8 g/cm$^3$), and porosity (8%), are within the range of the bulk (average 3.42 g/cm$^3$ s.d. 0.18 g/cm$^3$) and grain (3.72 g/cm$^3$ s.d. 0.12 g/cm$^3$) density and porosity (7.0% s.d. 4.90%) of other H chondrite falls reported in *Consolmagno et al. (2008)*. The magnetic susceptibility logarithm of 5.4 is also within the H chondrite falls value (average 5.32 s.d. 0.10) reported in *Rochette et al. (2003)*.

The total relative area (relative part) of the Mössbauer spectral components associated with Fe-Ni-Co alloy is significantly (~53 %) larger than this value found in the range ~10–20 % in the Mössbauer spectra of some other H ordinary chondrites by *Oshtrakh et al. (2008)* and *Maksimova et al. (2017)*. The absorption effect of the overlapped magnetic sextets related to Fe-Ni-Co alloy is smaller than that for troilite magnetic sextet in the abovementioned studies. In the Annama case the absorption effect for troilite magnetic sextet is about 3 times smaller than that for the total main overlapped Fe-Ni-Co alloy sextets. This fact indicates a larger content of Fe-Ni-Co alloy in the 88 mg Annama II sample used for the Mössbauer studies. The larger Fe-Ni-Co alloy content in the studied sample may be related to inhomogeneous (locally enriched) content of these phases in Annama source material at a ~10 mg scale or it may be a result of another composition of iron-bearing phases. We cannot distinguish among these two possible explanations and thus, further investigations are required.

The comparison of the total metal content derived from metallography (15 – 20 wt%) to the average value of 18.02 wt% found in other H chondrites (*McSween et al. 1991*) as well as abovementioned magnetic susceptibility value indicates that the bulk metal content of Annama is similar to other H chondrites.

We can also compare the ratios of $Fe^{2+}$ parts in the M1 and M2 sites in both olivine and orthopyroxene using XRD and Mössbauer data. The ratios $X_{Fe}^{M1}/X_{Fe}^{M2}$ obtained from XRD are 1.2 and 0.1 for olivine and orthopyroxene, respectively, while the relative area ratios $A_{M1}/A_{M2}$ in the Mössbauer spectrum are 1.4 and 0.2 for the same silicates, i.e. results obtained from the two methods are in agreement.

### *Cosmic ray exposure and thermal history*

Information on the cosmic ray exposure (CRE) age of Annama, its pre-entry size and the depth of the two Annama meteorite fragments (I and II) within the meteoroid can be inferred from the cosmogenic radionuclide and noble gas data. The He, Ne, Ar content of Annama can be explained by two major components: *cosmogenic noble gases* from spallation reactions induced by cosmic rays, and *radiogenic noble gases* from the decay of $^{40}K$ and $^{238}U$, $^{235}U$ and $^{232}Th$. No trapped (e.g., solar-wind derived) noble gases are present for He and Ne, but a small, trapped argon component is present, as indicated by a $^{36}Ar/^{38}Ar$ ratio higher than the cosmogenic value of 0.65 (Table 3). The radiogenic $^{4}He$ and $^{40}Ar$ concentrations provide information on the thermal history of the Annama meteorite on its parent body.

#### Cosmogenic radionuclides

The production rates of cosmogenic radionuclides in meteorite samples are mainly a function of the primary GCR (galactic cosmic ray) flux and the shielding conditions of the sample, i.e. the pre-entry size of the meteoroid and depth of the sample with the meteoroid. Activities of short-lived radionuclides, with half-lives less than the orbital period, integrate the production of these radionuclides over the last segment of the orbit and solar cycle, while activities of long-lived radionuclides (such as $^{10}Be$, $^{26}Al$ and $^{36}Cl$) integrate the production over many orbits and many solar cycles.

The $^{22}Na/^{26}Al$ activity ratio in stony meteorite falls is a useful indicator for the GCR flux in the last 3 – 5 years before the fall, as the $^{22}Na/^{26}Al$ production ratio is relatively independent of shielding conditions (*Evans et al. 1982*, *Bhandari et al. 2002*), so the measured ratio varies as a function of the $^{22}Na$ production rate over the past 3 – 5 years. The measured $^{22}Na/^{26}Al$ ratio of 1.7 ±0.2 in Annama is similar to the average ratio of 1.5 for H chondrite falls (*Evans et al. 1982*, *Bhandari et al. 2002*). The fall of the Annama H5 ordinary chondrite occurred in the middle of the solar cycle 24 as indicated by the Bartol Research Institute neutron monitor data (http://neutronm.bartol.udel.edu/, retrieved November 10, 2016). Since solar cycle 24 shows a relatively small maximum in solar activity and was preceded by a long period of low solar activity, the slightly elevated $^{22}Na/^{26}Al$ ratio indicates that the average GCR flux in 2008 – 2013 was <10% higher than the long-time (million year) average. It is therefore safe to assume that radionuclides with shorter half-lives than $^{22}Na$ experienced an average GCR flux very similar to the long-time average. Since the CRE age of Annama is ~30 Ma, as will be shown below, all radionuclides were saturated at the time of fall, and the measured activities thus represent the production rates, which can be compared to model calculations for meteorites of different size.

To derive the pre-entry size and depth of Annama II, we mainly used the $^{60}Co$ (5.27 a), $^{54}Mn$ (312 d), $^{22}Na$ (2.6 a) and $^{26}Al$ (0.705 Ma) activities. Of these four radionuclides, the production rate of $^{60}Co$ is the most sensitive to the size and depth of the meteorite, since this nuclide is produced from $^{59}Co$ by capture of low-energy thermal neutrons, which only reach significant fluxes in objects large enough to attenuate the secondary neutrons to thermal energies. As the $^{60}Co$ concentration of 7 dpm/kg in Annama II is rather low compared to maximum values of ~200 dpm/kg in objects with radii of 45 – 65 cm, the specimen was either from a small meteoroid (<20 cm radius) or close to the surface of meteoroid with radius larger than 20 cm (*Eberhardt et al. 1963*, *Spergel et al. 1986*). The measured $^{22}Na$ activity was compared to the calculations of *Bhandari et al. (1993)* for H chondrites, yielding a radius of 10 – 20 cm if the sample was from the center of the meteoroid, or a bigger meteoroid (20 – 100 cm) if it came from the surface. The $^{54}Mn$ activity was normalized to the concentration of its main target Fe (27 wt%) and compared to the calculations of *Kohman and Bender (1967)* giving a range for the radius of 10 – 20 cm, in case it came from the central part, or 20 – 100 cm, in case it came from anywhere in the meteoroid. Finally, the low $^{56}Co$ activity indicates that the sample did not originate from the outer 1 – 2 cm, where elevated $^{56}Co$ from production by solar cosmic rays (SCR) is expected (e.g. *Evans et al.1987*, *Nishiizumi et al. 2014*). Comparison of the $^{26}Al$ activity in Annama II with calculations of *Leya and Masarik (2009)* gives a similar size and depth range. Short-lived radionuclide and $^{26}Al$ from Annama II thus give us two possible scenarios – an origin in the center of a small (10 – 20 cm radius) meteoroid, or a near-surface origin (5 – 15 cm depth) on a larger meteoroid (up to ~1 m radius).

For Annama I, we only measured the long-lived radionuclides, $^{10}Be$, $^{26}Al$ and $^{36}Cl$. The $^{26}Al$ concentration of 73 ±3 dpm/kg in Annama I is significantly higher than the value of 54 ±5 dpm/kg measured in Annama II. Although these two values were determined with different methods (AMS vs. GRS), the two methods generally agree within 10 – 15% for the same sample, so the 35% difference in $^{26}Al$ between the two samples is real and is most likely due to a significant difference in shielding depth of the two Annama fragments within the pre-entry meteoroid. Based on the model of *Leya and Masarik (2009),* the measured $^{26}Al$ activity in Annama I is within error consistent with calculated production rates in or near the center of objects with radii of 40 – 50 cm (Fig. 11a), i.e. slightly larger than the size of 30 – 35 cm estimated from the fireball observations. In this scenario, Annama II comes from a depth of ~5 cm, which is consistent with the short-lived radionuclide results discussed above.

Most of the $^{36}$Cl in stone meteorites is produced from spallation reactions on Ca and Fe, with minor contributions from K, Ti, Cr, Mn and Ni (*Leya and Masarik 2009*). However, in objects large enough (R >20 cm) to attenuate the secondary neutrons to thermal energies, $^{36}$Cl is also produced from neutron capture on $^{35}$Cl. The $^{36}$Cl concentration of 9.8 dpm/kg in the stone fraction of Annama I is ~30% higher than the maximum $^{36}$Cl production rate of ~7.5 dpm/kg for spallation reactions alone (Fig. 11b). The elevated $^{36}$Cl indicates a significant contribution (2.5 ±0.5 dpm/kg) of neutron-capture-produced $^{36}$Cl. Assuming an average Cl concentration of 80 ppm for H chondrites, this yields a $^{36}$Cl production rate of ~30 atoms/min/g Cl. This neutron-capture $^{36}$Cl component corresponds to a $^{60}$Co activity of ~50 dpm/kg (*Spergel et al., 1986*), i.e. much higher than the neutron-capture $^{60}$Co activity of 7 dpm/kg in Annama II. This is consistent with the pre-entry radius of >30 cm and confirms that Annama I came from a deeper position than Annama II.

The $^{10}$Be concentration of 18.9 dpm/kg is ~10% lower than expected for irradiation in the center of an object with radius of 40 – 50 cm (Fig. 11c), a shielding condition that was inferred from the high $^{26}$Al concentration in Annama II. However, the measured $^{10}$Be concentration is consistent with irradiation closer to the surface of a larger object (R~65 cm). The disagreement between shielding conditions derived from the $^{26}$Al and $^{10}$Be data is also evident from the measured $^{26}$Al/$^{10}$Be ratio of ~4.0, which is higher than the production ratio of 3.0 – 3.5 for medium to large objects. This elevated $^{26}$Al/$^{10}$Be ratio could indicate that Annama experienced a recent break-up event 3 – 5 Ma ago, which led to higher production rates in the last stage and left $^{10}$Be slightly below saturation for the new shielding conditions. This event may be responsible for reducing the pre-entry size of Annama to 30 – 35 cm radius inferred from atmospheric entry modeling. Such a complex exposure scenario seems consistent with the noble gas data in Annama, which – as will be discussed below – also indicate higher shielding conditions for most of its exposure history than those inferred from fireball observations. To constrain the details of this complex exposure history would require more detailed analyses, which are beyond the scope of this initial study of Annama.

## Cosmogenic noble gases

The cosmogenic $^{22}$Ne/$^{21}$Ne ratio is a characteristic shielding indicator for small to medium sized meteoroids (<65 cm radius), where it varies from approximately 1.25 near the surface of very small (chondritic) meteoroids to approximately 1.06 in the center of an object with 65 cm radius (*Leya and Masarik 2009*). The average measured $^{22}$Ne/$^{21}$Ne ratio in Annama Ia + Ib (1.058 ±0.003) is at the low end of this range. In agreement with the radionuclide findings presented above, this suggests that (most of) the cosmogenic Ne in Annama was produced in a meteoroid with a radius of >65 cm and (at H chondrite density) a mass of >4000 kg. This is larger than the calculated pre-entry mass of Annama based on the fireball data analysis (*Trigo-Rodríguez et al. 2015, Lyytinen and Gritsevich 2016*), suggesting that the meteorite had a complex exposure history, i.e. was exposed to cosmic radiation under more than one shielding condition. The $^{3}$He/$^{21}$Ne ratio of Annama Ia + Ib is 3.65, which is a bit lower than expected for a meteorite with $^{22}$Ne/$^{21}$Ne = 1.058 (based both on the *Leya and Masarik (2009)* model and the "Bern-line"; *Eberhardt et al. 1966*), suggesting He loss on the order of ~15%. Using the $^{22}$Ne/$^{21}$Ne ratio as a shielding parameter as suggested by *Leya and Masarik (2009)*, we derive a $^{21}$Ne production rate of 4.1 x $10^{-9}$ $cm^3$ STP/gMa and a cosmic ray exposure age of 32 Ma. If we use the radius and shielding depth suggested by $^{22}$Ne/$^{21}$Ne = 1.058 (R = 65 cm, D = 46 cm) to directly derive production rates from the model, the $^{3}$He, $^{21}$Ne, and $^{38}$Ar exposure ages are 28, 34 and 27 Ma, respectively. Another method to determine the CRE age is to use the $^{26}$Al/$^{21}$Ne ratio. Since the production mechanism for $^{26}$Al and $^{21}$Ne is very similar, the activity of $^{26}$Al can be used to estimate the $^{21}$Ne production rate (e.g. *Graf et al. 1990*). Adopting a $^{26}$Al/$^{21}$Ne production rate ratio of ~0.35 at/at (*Leya and Masarik, 2009*), the measured $^{26}$Al/$^{21}$Ne ratio yields a CRE age of 32 Ma. The uncertainty on these ages is about 15% and is dominated by the typical uncertainties of the production rate model. Our preferred value is 30 ±4 Ma, which should however be considered to be a lower limit given the possibility of a complex exposure history with a long first-stage exposure under high shielding. Annama clearly does not belong to the main cluster of H chondrites with CRE ages of 7 – 8 Ma (*Marti and Graf 1992*).

Radiogenic noble gases

Using a cosmogenic $^{4}$He/$^{3}$He ratio of 5.8 corresponding to the shielding conditions inferred from the cosmogenic $^{22}$Ne/$^{21}$Ne ratio (*Leya and Masarik 2009*), we derive the non-cosmogenic $^{4}$He concentrations of Annama. If is the remaining $^{4}$He is interpreted as entirely radiogenic, from the decay of long-lived radionuclides $^{235}$U, $^{238}$U and $^{232}$Th present at chondritic abundances (13 ppb U and 43 ppb Th, *Wasson and Kallemeyn 1988*), the resulting nominal U,Th-He age is 2.7 Ga for Annama Ia + Ib (2.5 Ga for Annama Ir). When allowing for a 15% He loss as suggested by cosmogenic He, this value increases to 3.0 Ga. Similarly, a nominal K-Ar age can be determined from the measured $^{40}$Ar (which should be almost completely radiogenic) and a chondritic K abundance (780 ppm, *Wasson and Kallemeyn 1988*). This yields a retention age of 3.8 Ga (for all three fragments of Annama I analyzed) and suggests that Annama has never been heated to the point of Ar loss since the time of the Late Heavy Bombardment (see however Ar-Ar dating in next section suggesting a higher Ar-Ar age for Annama). Assuming conservatively that $^{36}Ar_{trapped}$ is entirely of atmospheric origin with an atmospheric $^{40}$Ar/$^{36}$Ar ratio of 296, we find that at most 7% of the measured $^{40}$Ar could be atmospheric.

The concentrations of U, Th and K, derived from the measured activities of the natural radionuclides $^{232}$Th and $^{238}$U as well as for $^{40}$K in the meteorite specimens are listed in

Table 4. Their values are well in accordance with the average concentrations given in *Wasson and Kallemeyn (1988)* for H chondrites.

*Ar-Ar dating*

The Ar-Ar plateau age of Annama of 4420 ±13 Ma at 675 – 780°C is about 80 Ma younger than observed for H5 chondrites with a reported shock stage S1 (~4500 Ma, *Trieloff et al. 2003*) which may indicate an impact-induced resetting of the K-Ar system. This might be supported by the degassing pattern of $^{39}$Ar which dominantly was released at lower temperatures (about 92% of the total $^{39}$Ar was released up to 780°). This would be expected if a slight shock cracks the respective K carrier grains and thus changes the grain size distribution towards smaller effective diffusional lengths. However, the degassing characteristics alone is not a criteria for shock effects. We observe one major release peak of $^{39}$Ar, in opposite to some other hardly shocked H5 chondrites which display a second major degassing peak at elevated temperatures >1000°C ( *Trieloff et al. 2003*). Annama is classified with a shock stage S2, but it contains thin veins of impact melt which may point to a more severe thermal event potentially related to our younger age. It is thus possible that the low temperature plateau age of 4420 ±13 Ma represents the time of this thermal event. From our Ar-Ar analyses we only obtained a maximum cosmic ray exposure age of about 40 Ma calculated from Ca-derived spallogenic $^{38}$Ar, which is however, in reasonable agreement with our noble gas data.

## Conclusions

Based on mineralogy and physical properties, Annama is a typical H chondrite. It also has a high Ar-Ar age of 4.4 Ga. Its cosmic ray exposure history is, however, atypical. Annama is not part of the large group of H chondrites with a prominent 7 – 8 Ma peak in the exposure age histograms. Instead, its exposure age is within uncertainty of a smaller peak at 30 ±4 Ma. The combination of noble gases and radionuclides also seem to indicate a complex exposure history. The results from short-lived radionuclides are compatible with an atmospheric pre-entry radius of 30 – 40 cm. However, based on noble gas and cosmogenic radionuclide data, Annama must have been part of a larger body (radius >65 cm) for a large part of its cosmic ray exposure history. The $^{10}$Be concentration indicates a recent (3 – 5 Ma) breakup which may be responsible for the Annama parent body size reduction to 30 – 35 cm pre-entry radius.

## Acknowledgements

The work was supported by the Academy of Finland projects nos. 257487, 260027, and 293975. Institute of Geology, the Czech Academy of Sciences is supported by Ministry of Education, Youth and Sports project no. RVO67985831. M. G. acknowledges support from the ERC Advanced Grant no. 320773, and the Russian Foundation for Basic Research (RFBR), projects nos. 14-08-00204, 16-05-00004, and 16-07-01072. Research at the Ural Federal University is supported in part by the Ministry of Education and Science of the Russian Federation (basic financing for the projects nos. 2085 (M. I. O.) and 1514 (V. I. G.) in 2016 and financing for the project no. 1959 (M.I.O., A.A.M., A.V.C. and V.A.S.) in 2017) and the Act 211 of the Government of the Russian Federation, agreement no. 02.A03.21.0006. Contribution to the study from A. A. M. was funded by the RFBR according to the research project no. 16-32-00151 mol_a. M. M. was supported by an SNSF Ambizione grant (no. PZ00P2_154874). J. H. and M. T. acknowledge support by Klaus Tschira foundation. R. R. M. thanks support from Department of Space, Govt. of India. J.M.T-R, and C.E.M-C received financial support from the Spanish Ministry (research project AYA2015-67175-P). The authors acknowledge being a part of the network supported by the COST Action TD1403 "Big Data Era in Sky and Earth Observation" and TD1308 "ORIGINS".

Table 1. Measured concentrations of major elements (in wt%) and cosmogenic radionuclides (in dpm/kg) in stone fraction of the specimen from Annama I. O and Si concentrations were estimated from other elements and bulk composition of H chondrites (*Wasson and Kallemeyn 1988*).

| Element | Conc. (wt%) in stone |
|---|---|
| O | 43.0 |
| Na | 0.9 |
| Mg | 16.5 |
| Al | 1.3 |
| Si | 20.5 |
| K | 0.098 |
| Ca | 1.53 |
| Ti | 0.07 |
| Mn | 0.26 |
| Fe | 14.1 |
| Co | 0.01 |
| Ni | 0.36 |
| | |
| Radionuclides | Conc. (dpm/kg) |
| $^{10}$Be | 22.3 ±0.3 |
| $^{26}$Al | 89.4 ±2.5 |
| $^{36}$Cl | 9.8 ±0.2 |

Table 2. Mass-normalized activities (corrected to the time of fall) of cosmogenic radionuclides measured on a specimen from Annama II by non-destructive gamma ray spectroscopy. Errors include a 1σ uncertainty of ~10% in the detector efficiency calibration.

| Nuclide | Half-life | Activity (dpm/kg) |
|---|---|---|
| $^{7}$Be | 53.2 d | 200 ± 30 |
| $^{58}$Co | 70.9 d | 12 ± 2 |
| $^{56}$Co | 77.2 d | 4 ± 2 |
| $^{46}$Sc | 83.8 d | 9 ± 2 |
| $^{57}$Co | 271.8 d | 13 ± 2 |
| $^{54}$Mn | 312.3 d | 111 ± 11 |
| $^{22}$Na | 2.60 y | 91.5 ± 7.2 |
| $^{60}$Co | 5.27 y | 7 ± 1 |
| $^{44}$Ti | 60 y | <3.7 |
| $^{26}$Al | 7.05 x $10^{5}$ y | 54.3 ±4.6 |

Table 3: He, Ne, Ar in Annama Ia + Ib. All concentrations given in $10^{-8}$ $cm^3$ STP/g, all ages in Ma.

| | Annama Ia | Annama Ib | Total |
|---|---|---|---|
| Mass (g) | 0.114 | 0.0537 | 0.168 |
| $^{3}He/^{4}He$ | 0.0384 | 0.0402 | 0.0390 |
| $^{4}He$ | 1240 | 1190 | 1220 |
| $^{4}He_{rad}$ | 965 | 909 | 947 |
| $^{20}Ne/^{22}Ne$ | 0.854 | 0.849 | 0.852 |
| $^{21}Ne/^{22}Ne$ | 0.946 | 0.943 | 0.945 |
| $^{22}Ne$ | 13.6 | 14.2 | 13.8 |
| $^{36}Ar/^{38}Ar$ | 1.14 | 1.17 | 1.15 |
| $^{40}Ar/^{36}Ar$ | 2420 | 2530 | 2450 |
| $^{36}Ar$ | 1.58 | 1.64 | 1.60 |
| $^{40}Ar_{rad}$ | 3810 | 4150 | 3920 |
| $^{3}He/^{21}Ne_{cos}$ | 3.71 | 3.56 | 3.66 |
| $^{22}Ne/^{21}Ne_{cos}$ | 1.057 | 1.060 | 1.058 |
| T3 | 34.0 | 34.0 | 34.0 |
| T21 | 31.4 | 32.6 | 31.8 |
| T38 | 26.9 | 27.1 | 26.9 |
| R4 | 2740 | 2620 | 2710 |
| R40 | 3750 | 3890 | 3800 |

Table 4. Concentration of primordial radionuclides measured on a specimen from Annama II by non-destructive gamma ray spectroscopy. Errors include a 1$\sigma$ uncertainty of ~10% in the detector efficiency calibration.

| Nuclide | Concentration |
|---|---|
| $^{238}U$ | 11 ±1 ng/g |
| $^{232}Th$ | 39 ±4 ng/g |
| $^{40}K$ | 800 ±80 µg/g |

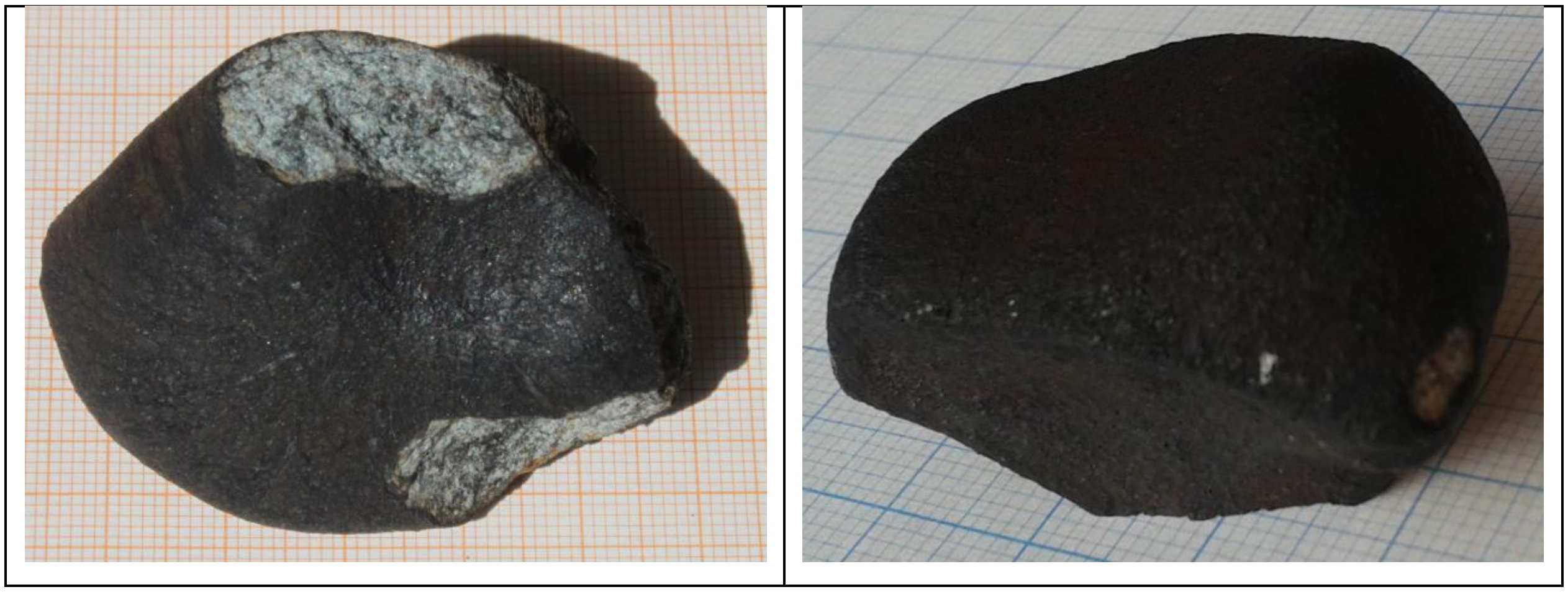

Fig. 1. 120.4 g Annama I (left) and 47.5 g Annama II (right) meteorite. The minor grid size in background is 1 mm.

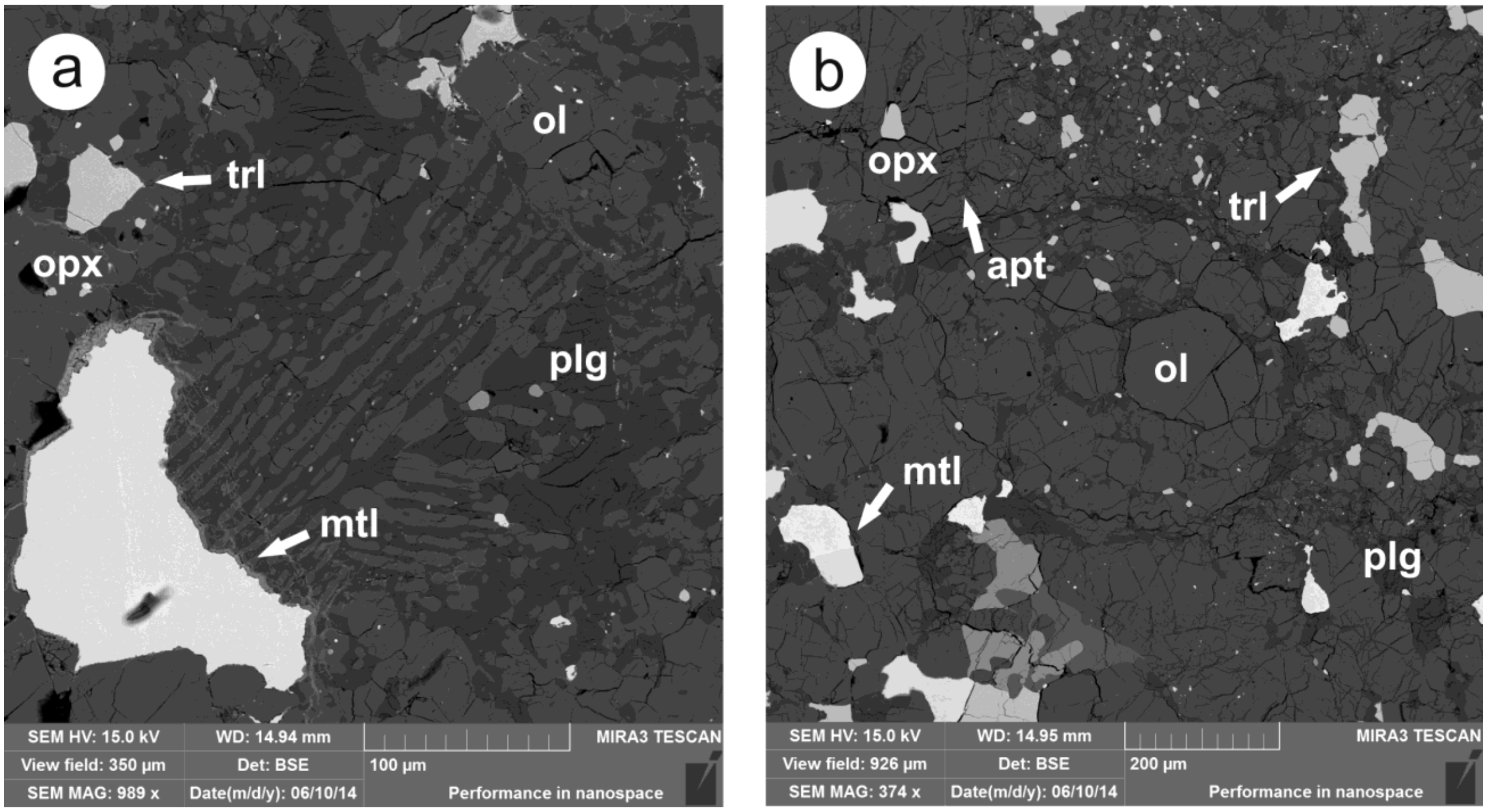

Fig. 2.SEM BSE (Scanning electron microscope – back-scattered electrons) image of two chondrules (apt – apatite, mtl – metal, ol – olivine, opx – orthopyroxene, pgl, – plagioclase, trl – troilite).

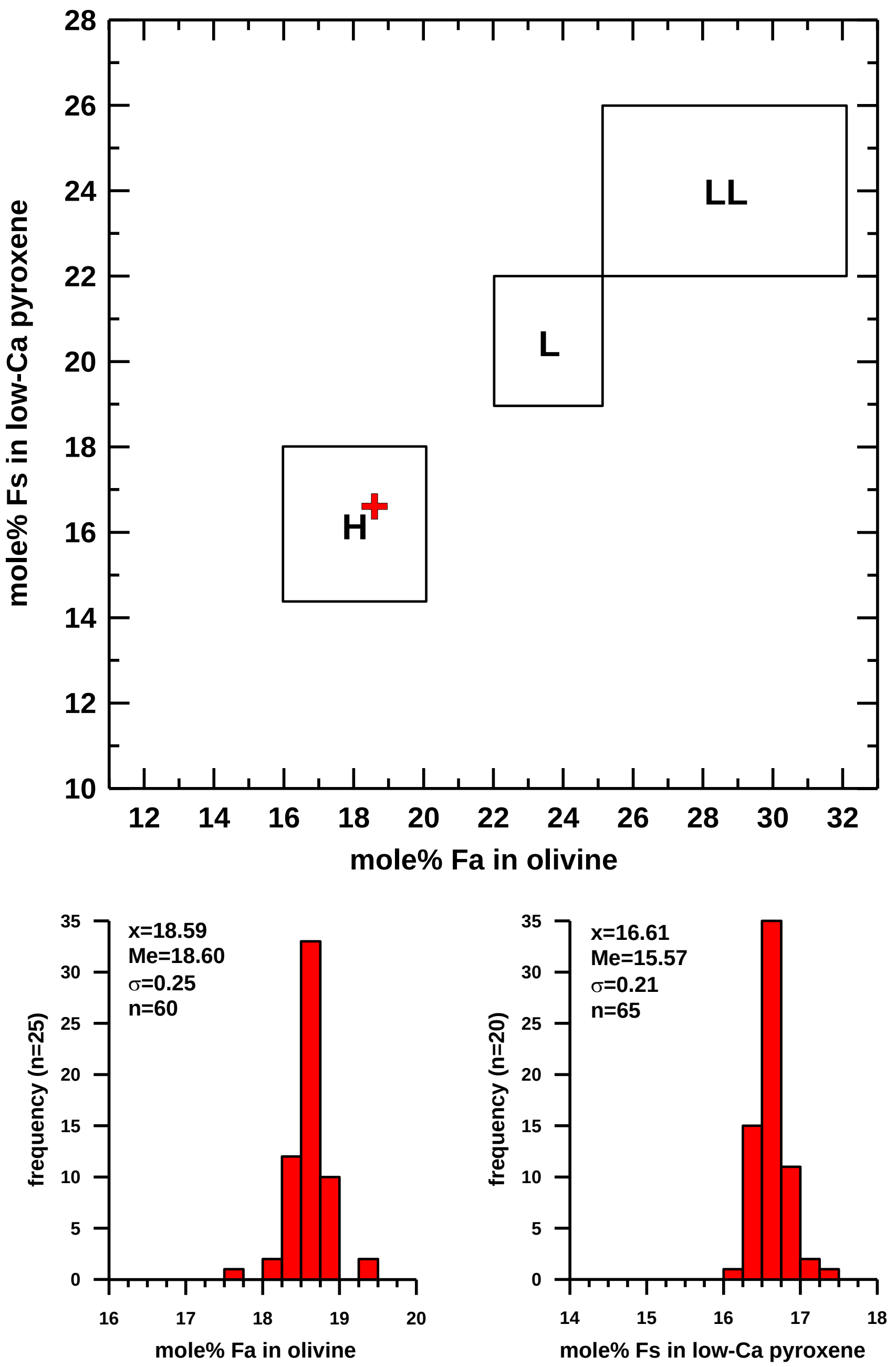

Fig. 3. Classification (up) on Annama (red cross) and histogram of olivine (down left) and low-Ca pyroxene (down right) compositions.

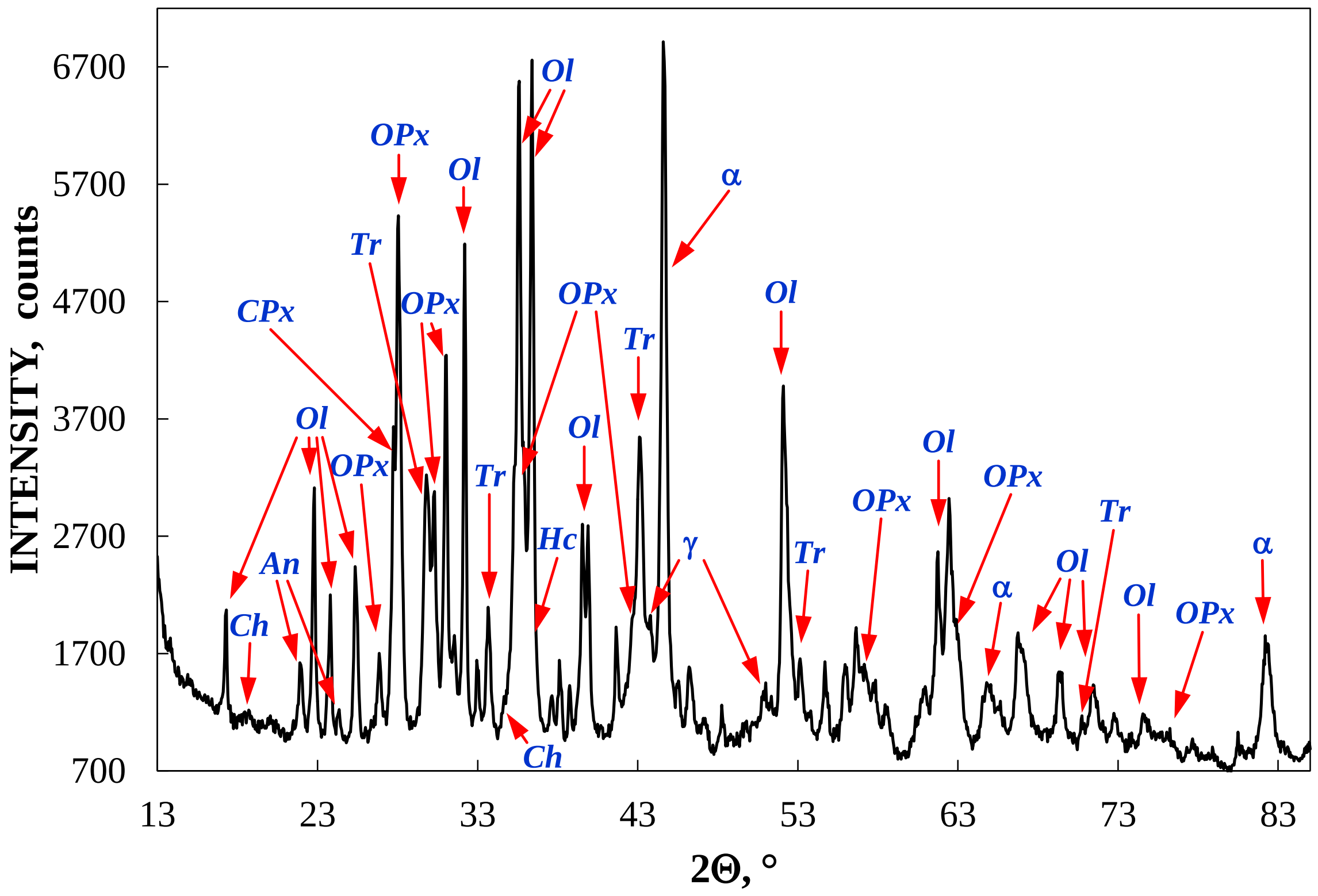

Fig. 4. XRD (X-Ray diffraction) pattern of Annama. Arrows indicate selected reflexes (Ol – olivine, An – anorthite, OPx – orthopyroxene, CPx – clinopyroxene, Tr – troilite, α – metallic α-phase, γ – metallic γ-phase, Ch – chromite, Hc – hercynite).

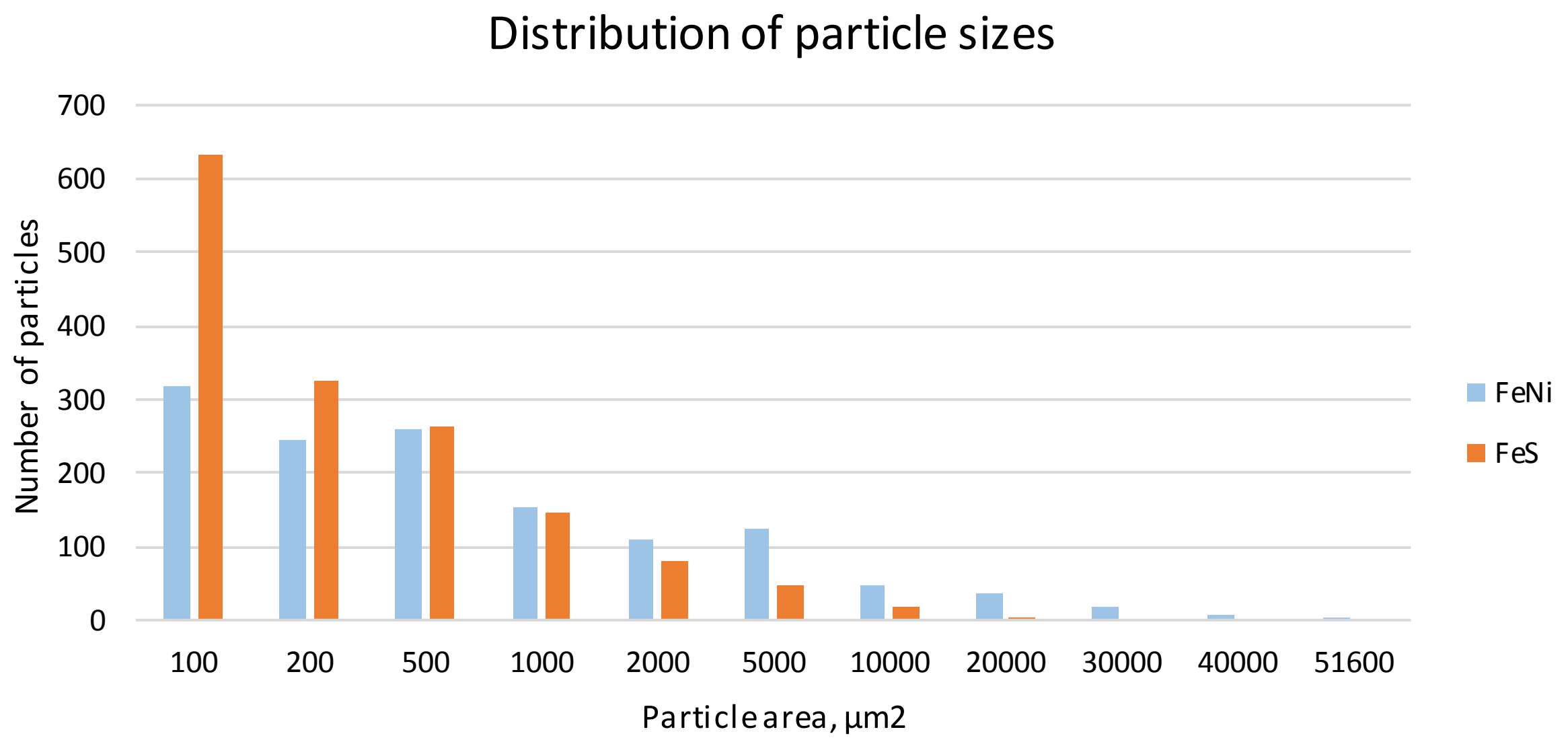

Fig. 5. Size distribution of metal and troilite particles in polished section of Annama I.

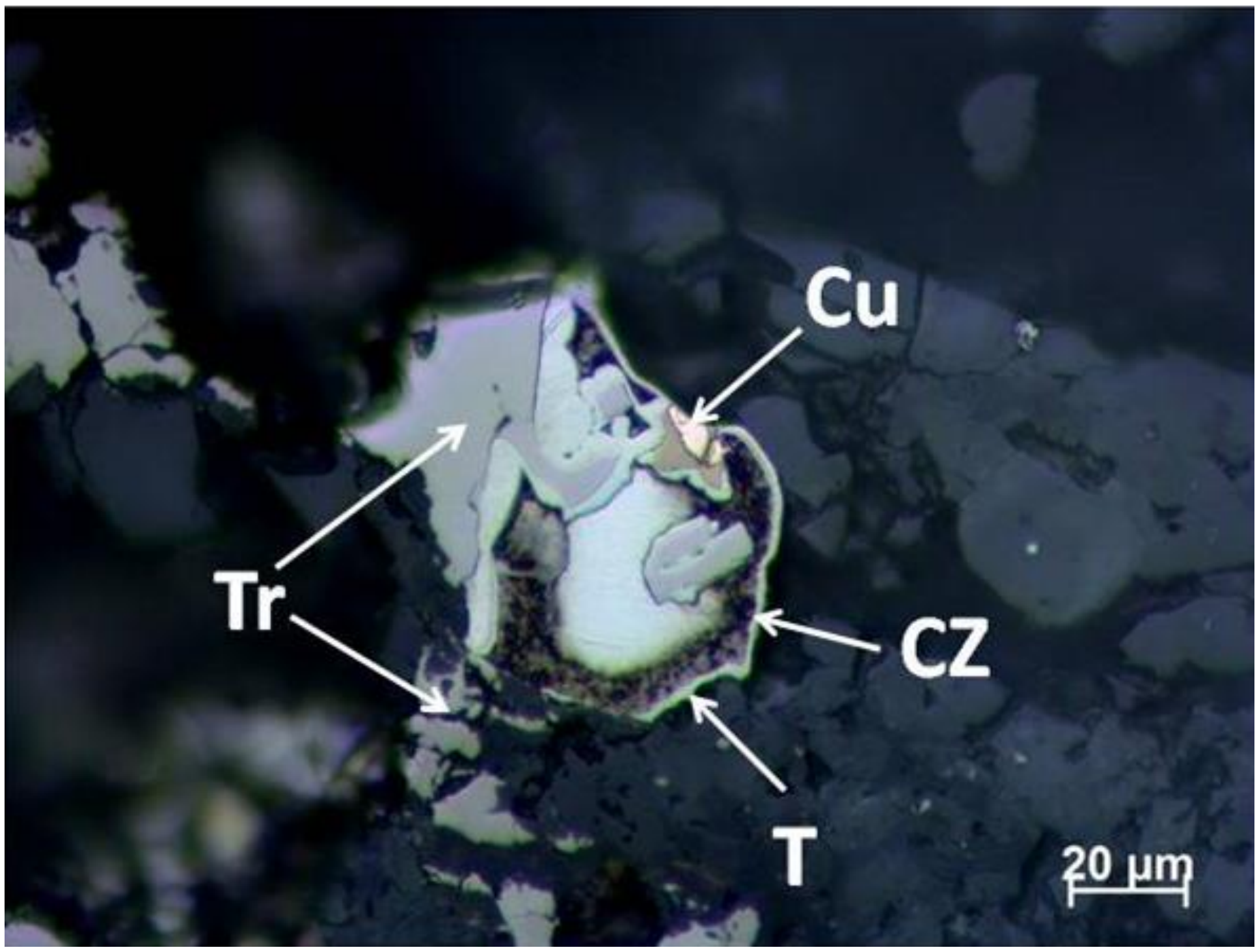

Fig. 6. Microphotograph of zoned metal particle with tetrateanite (T), cloudy zone (CZ), troilite (Tr) and copper crystal (Cu).

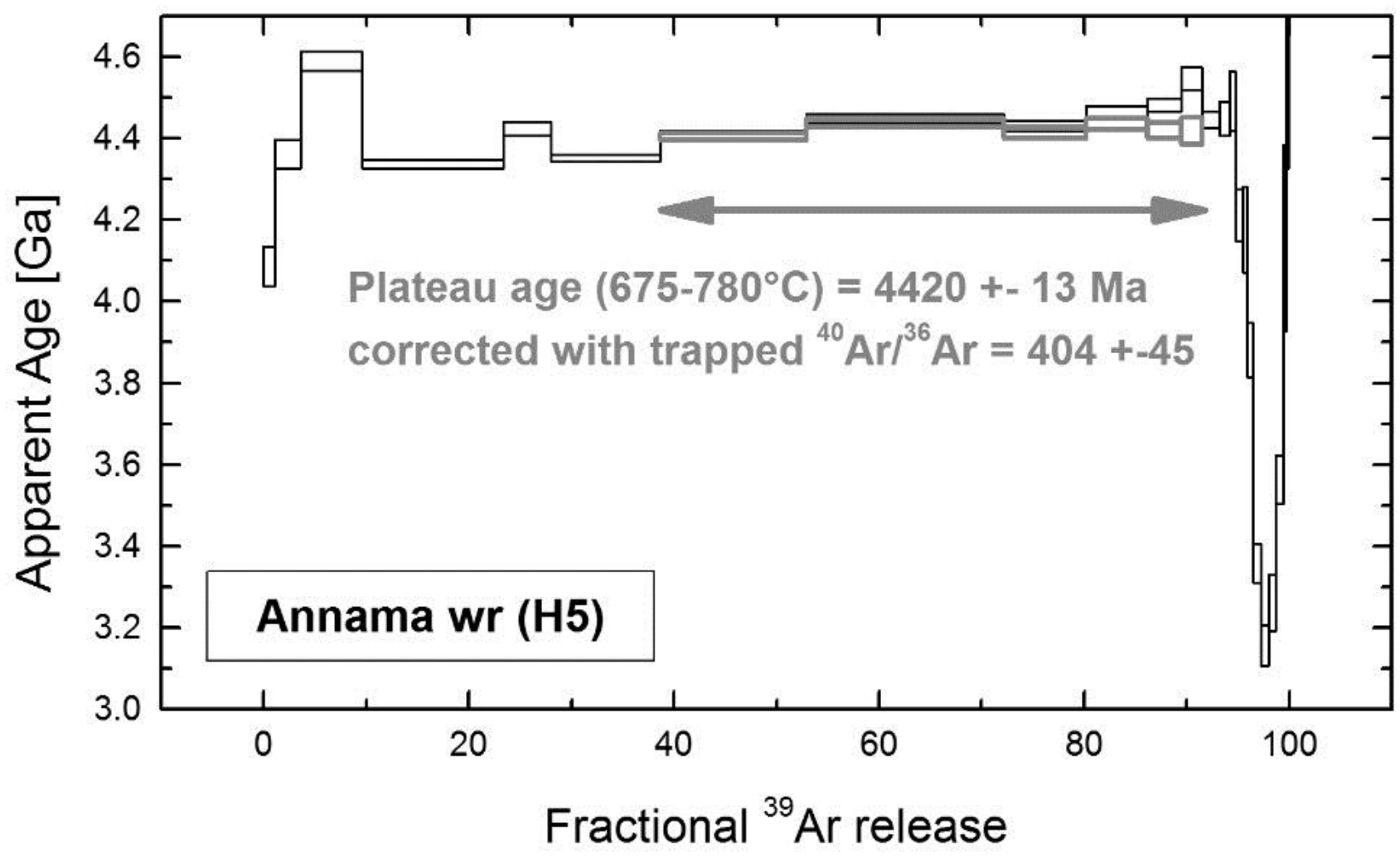

Fig. 7. Argon age spectrum of Annama. Black spectrum represents apparent ages corrected with a trapped solar $^{40}Ar/^{36}Ar$ composition of 1 ±1. A correction with a trapped $^{40}Ar/^{36}Ar$ ratio of 404 ±45 resulted in a plateau age of 4420 ±13 Ma (grey color) which is more appropriate for isochron extractions of 675 – 780°C. Uncertainties are 1 s.d.

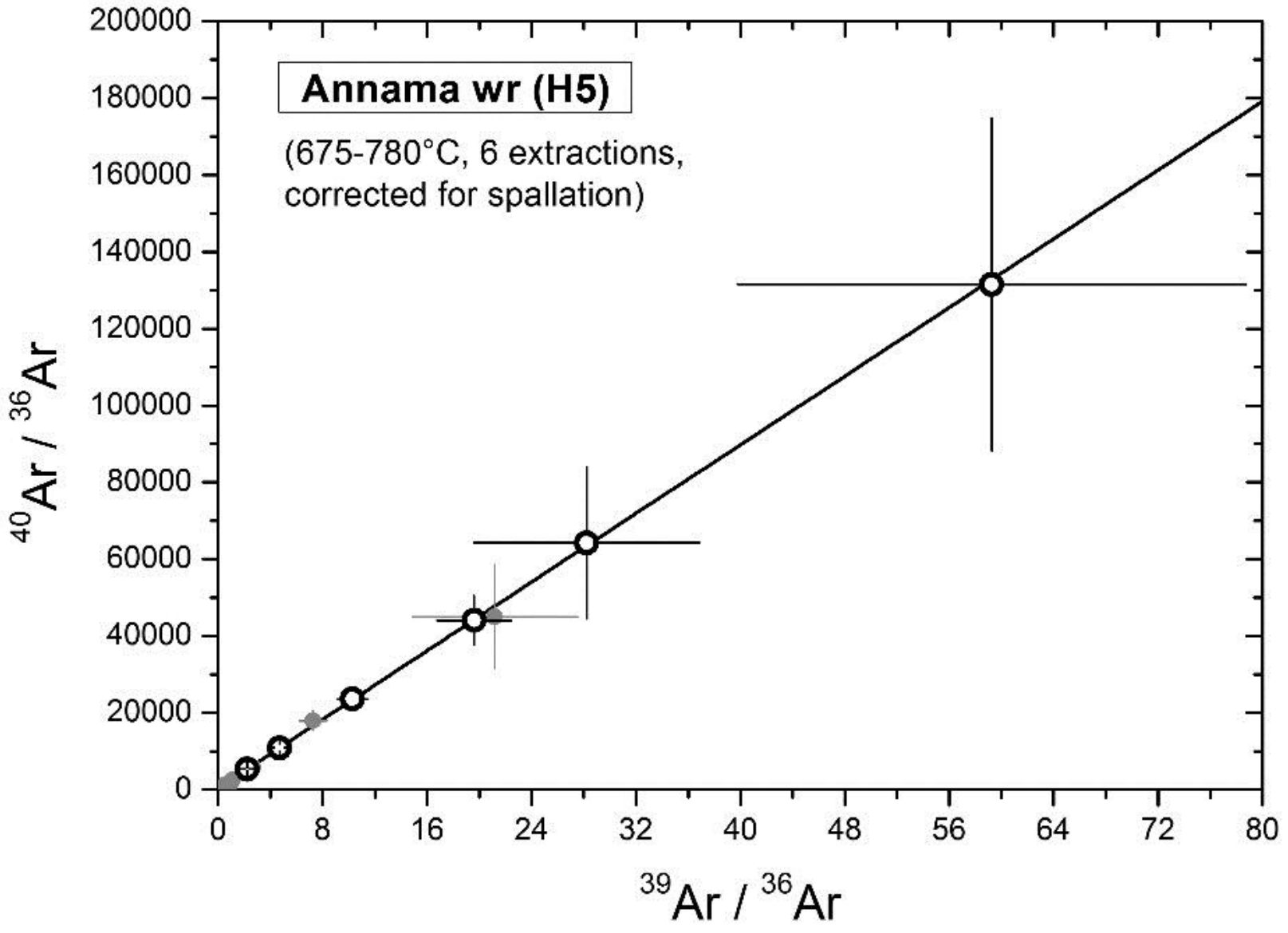

Fig. 8. Normal isochron diagram showing (675 – 780°C, N=6) isochron extractions (open black circles) and other data (grey dots). Results of regression can be found in Table S1 in Supporting Information. Error bars are 1 s.d. uncertainties.

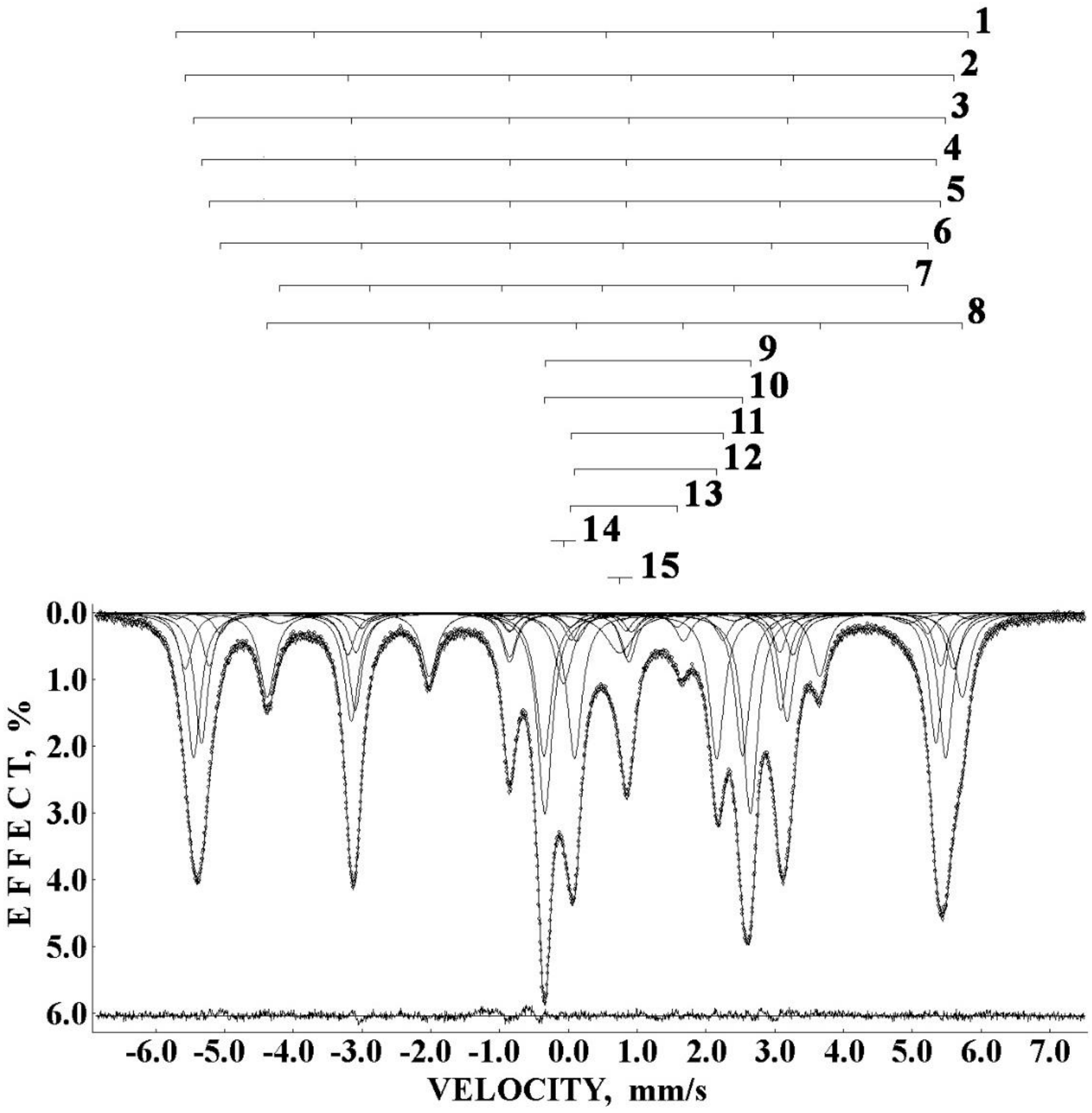

Fig.9. Mössbauer spectrum of Annama at 295 K. Indicated components are the result of the best fit described in the text. Differential spectrum is shown at the bottom of the figure.

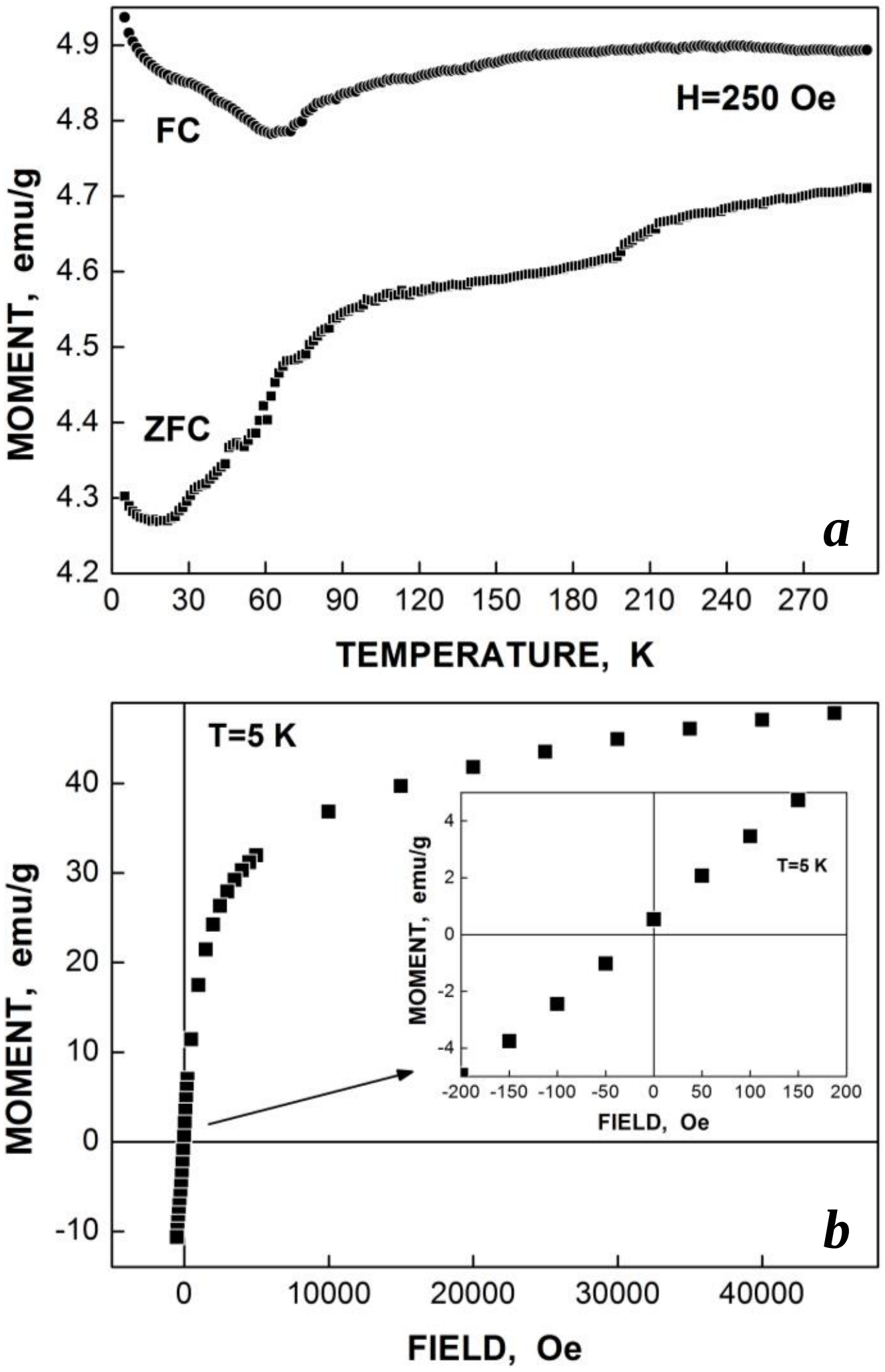

Fig. 10. Field cooled (FC) and zero field cooled (ZFC) curves (a) and isothermal magnetization (M( H)) curve (b) of Annama.

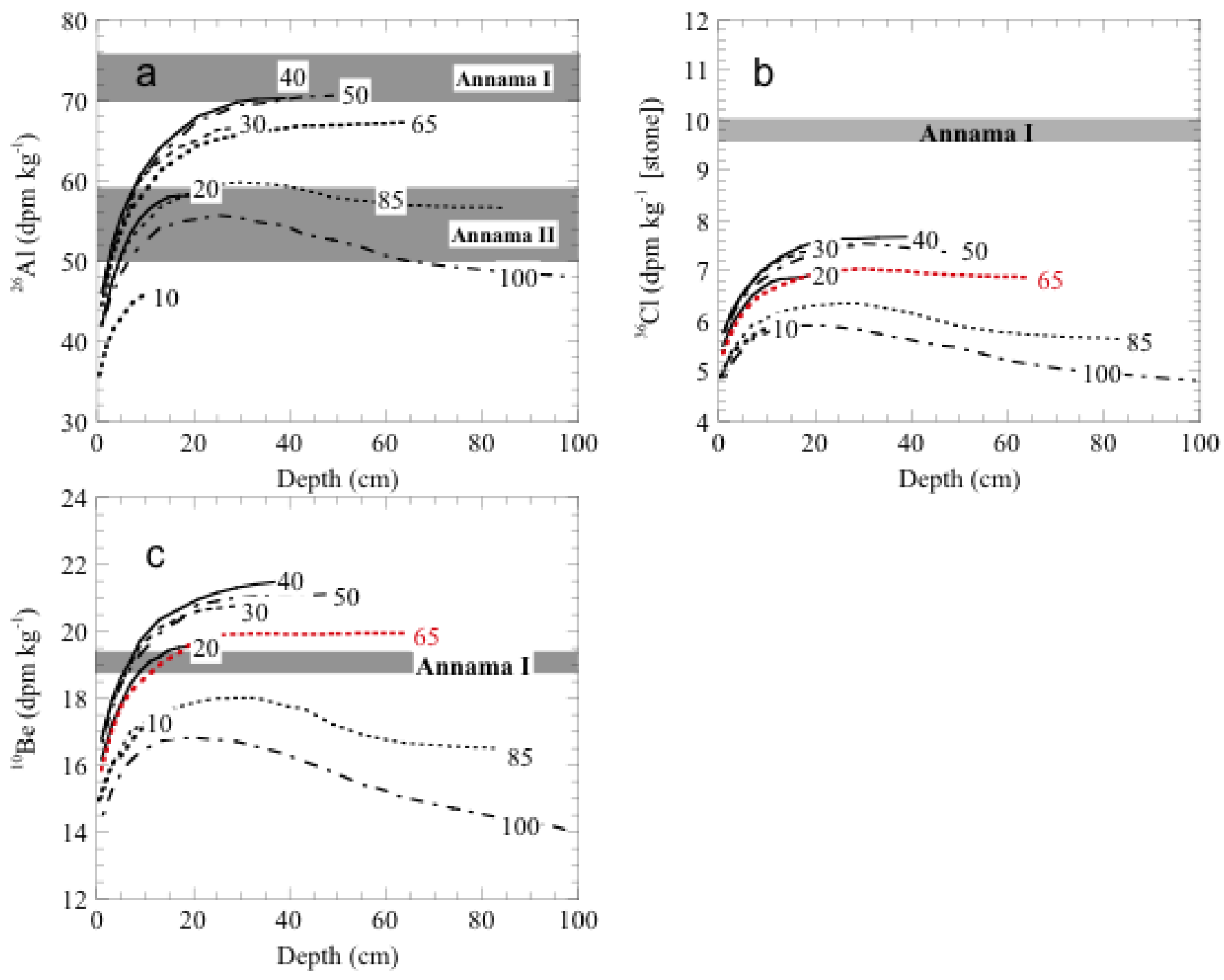

Fig. 11. Comparison of measured activities of $^{26}$Al (a), $^{36}$Cl (b) and $^{10}$Be (c) activities in Annama with calculated production rates (from *Leya and Masarik 2009*) for H chondrites with radii of 10 – 100 cm. For $^{26}$Al both the GRS (Gamma-ray spectroscopy) result for Annama II and AMS (Accelerator mass spectrometry) result for Ann ama I are shown. Note that the measured $^{36}$Cl concentration in Annama I is ~30% higher than calculated $^{36}$Cl production rates by spallation reactions alone, indicating a significant contribution of neutron-capture $^{36}$Cl.

**Supporting Information**

Table S1 is provided upon request (tomas.kohout@helsinki.fi) as an xlsx file .

Table S2. Concentrations and isotopic ratios of He, Ne, Ar, Kr and nitrogen in Annama-Ir. He, Ne, Ar concentrations are in $10^{-8}$ $cm^3$STP/g, Kr concentration is in $10^{-12}$ $cm^3$STP/g. Values are blank* corrected.

| Temp. °C | $^{4}He$ | $^{22}Ne$ | $^{36}Ar$ | $^{84}Kr$ | $^{3}He/^{4}He$ | $^{20}Ne/^{22}Ne$ | $^{21}Ne/^{22}Ne$ | $^{38}Ar/^{36}Ar$ | $^{40}Ar/^{36}Ar$ | $N_2$(ppm) | $\delta^{15}N$ (‰) | $^{82}Kr/^{84}Kr$ | $^{83}Kr/^{84}Kr$ | 8 |
|---|---|---|---|---|---|---|---|---|---|---|---|---|---|---|
| 600 | 272 | 0.33 | 0.05 | b.l. | 0.0252 ±0.0011 | 0.860 ±0.001 | 0.898 ±0.001 | 0.879 ±0.001 | 9069 ±2 | 1.1 | -30.22 ±0.07 | b.l. | b.l. | |
| 900 | 674 | 4.38 | 0.35 | 15.6 | 0.0351 ±0.0015 | 0.778 ±0.001 | 0.955 ±0.001 | 0.914 ±0.001 | 8978 ±29 | 36.5 | 4.89 ±0.13 | 0.2325 ±0.0005 | 0.2349 ±0.0003 | |
| 1200 | 81 | 8.66 | 0.86 | 20.3 | 0.0507 ±0.0022 | 0.818 ±0.001 | 0.968 ±0.001 | 0.882 ±0.001 | 193.4 ±0.1 | 1.3 | 28.25 ±0.04 | 0.2529 ±0.0003 | 0.2673 ±0.0011 | |
| 1700 | 5 | 4.58 | 0.98 | 46.1 | 0.1701 ±0.0074 | 0.796 ±0.001 | 0.987 ±0.001 | 0.972 ±0.001 | 39.8 ±0.1 | 0.3 | -43.48 ±0.15 | 0.2206 ±0.0002 | 0.2247 ±0.0002 | |
| Total | 1032 | 17.9 | 2.25 | 82.0 | 0.0344 ±0.0015 | 0.803 ±0.001 | 0.968 ±0.001 | 0.926 ±0.001 | 1710 ±4 | 39.2 | 4.37 ±0.13 | 0.2309 ±0.0002 | 0.2372 ±0.0005 | |

b. l. = blank level, * blank at 1700°C are, in $cm^3$STP for noble gases, as follows: $^{4}He = 2.23 \times 10^{-9}$, $^{22}Ne = 4.89 \times 10^{-11}$, $^{36}Ar = 3.39 \times 10^{-11}$, $^{84}Kr = 4.57 \times 10^{-13}$, and $N_2$ = 62 ng/V.

Table S3. Concentration and isotopic ratios of Xe in Annama Ir, Values are blank* corrected.

| Temp. °C | $^{132}$Xe $10^{-12}$ cm$^3$STP/g | $^{124}$Xe/$^{132}$Xe | $^{126}$Xe/$^{132}$Xe | $^{128}$Xe/$^{132}$Xe | $^{129}$Xe/$^{132}$Xe | $^{130}$Xe/$^{132}$Xe | $^{131}$Xe/$^{132}$Xe | $^{134}$Xe/$^{132}$Xe | $^{136}$Xe/$^{132}$Xe |
|---|---|---|---|---|---|---|---|---|---|
| 900 | 1.22 | 0.0058 | 0.0062 | 0.0873 | 1.145 | 0.1671 | 0.8329 | 0.3764 | 0.3105 |
| | | ±0.0001 | ±0.0004 | ±0.0002 | ±0.003 | ±0.0005 | ±0.0026 | ±0.0010 | ±0.0021 |
| 1200 | 25.8 | 0.0083 | 0.0107 | 0.0923 | 1.213 | 0.1679 | 0.8547 | 0.3841 | 0.3229 |
| | | ±0.0001 | ±0.0001 | ±0.0003 | ±0.003 | ±0.0003 | ±0.0034 | ±0.0006 | ±0.0003 |
| 1700 | 74.1 | 0.0053 | 0.0052 | 0.0834 | 1.120 | 0.1625 | 0.8241 | 0.3802 | 0.3123 |
| | | ±0.0001 | ±0.0001 | ±0.0001 | ±0.002 | ±0.0001 | ±0.0005 | ±0.0008 | ±0.0003 |
| Total | 112.1 | 0.0061 | 0.0066 | 0.0859 | 1.144 | 0.1642 | 0.8321 | 0.3807 | 0.3146 |
| | | ±0.0009 | ±0.0001 | ±0.0002 | ±0.002 | ±0.0002 | ±0.0013 | ±0.0007 | ±0.0005 |

*Blank at 1700°C is $^{132}$Xe = 1.66 x $10^{-15}$ cm$^3$STP.

**Table S4.** Mössbauer parameters for the main iron-bearing phases in a specimen from Annama I. Γ - line width, δ - isomer shift, $\Delta E_Q$ - quadrupole splitting (quadrupole shift for magnetically split spectra), $H_{eff}$ - magnetic hyperfine field, A - relative area of spectral components.

| Γ (mm/s) | δ (mm/s) | $\Delta E_Q$ (mm/s) | $H_{eff}$ (kOe) | A (%) | Component[a] |
|---|---|---|---|---|---|
| 0.233 ±0.014 | –0.164±0.009 | 0.420 ±0.018 | 356.2 ±0.7 | 0.69 | $\alpha_2$-Fe(Ni,Co) (1) |
| 0.257 ±0.014 | 0.022 ±0.007 | –0.012 ±0.007 | 345.5 ±0.5 | 10.23 | $\alpha_2$-Fe(Ni,Co) (2) |
| 0.236 ±0.014 | 0.015 ±0.007 | 0.012 ±0.007 | 338.2 ±0.5 | 17.06 | $\alpha$-Fe(Ni,Co) (3) |
| 0.233 ±0.014 | 0.004 ±0.007 | 0.015 ±0.007 | 330.4 ±0.5 | 15.09 | $\alpha$-Fe(Ni,Co) (4) |
| 0.233 ±0.014 | 0.055 ±0.007 | 0.140 ±0.009 | 328.0 ±0.5 | 4.50 | $\alpha$-Fe(Ni,Co) (5) |
| 0.233 ±0.014 | 0.044 ±0.007 | 0.131 ±0.013 | 315.6 ±0.6 | 1.07 | $\gamma$-Fe(Ni,Co) (6) |
| 0.401 ±0.021 | 0.061 ±0.008 | 0.606 ±0.015 | 283.9 ±0.7 | 2.05 | $\gamma$-FeNi (7) |
| 0.275 ±0.014 | 0.762 ±0.007 | Not determined | 313.8 ±0.5 | 11.42 | Troilite (8) |
| 0.260 ±0.014 | 1.150 ±0.007 | 2.992 ±0.007 | – | 12.51 | Olivine M1 (9) |
| 0.260 ±0.014 | 1.093 ±0.007 | 2.878 ±0.007 | – | 8.84 | Olivine M2 (10) |
| 0.260 ±0.014 | 1.139 ±0.007 | 2.200 ±0.014 | – | 1.97 | Pyroxene M1 (11) |
| 0.260 ±0.014 | 1.121 ±0.007 | 2.069 ±0.007 | – | 8.72 | Pyroxene M2 (12) |
| 0.260 ±0.014 | 0.852 ±0.007 | 1.465 ±0.007 | – | 0.87 | Hercynite (13) |
| 0.307 ±0.014 | –0.068 ±0.007 | – | – | 2.61 | $\gamma$-Fe(Ni,Co)$_{par}$ (14) |
| 0.498 ±0.014 | 0.748 ±0.007 | – | – | 2.35 | Chromite (15) |

[a]Numbers in parentheses correspond to the numbers of the spectral components in Fig. 9.

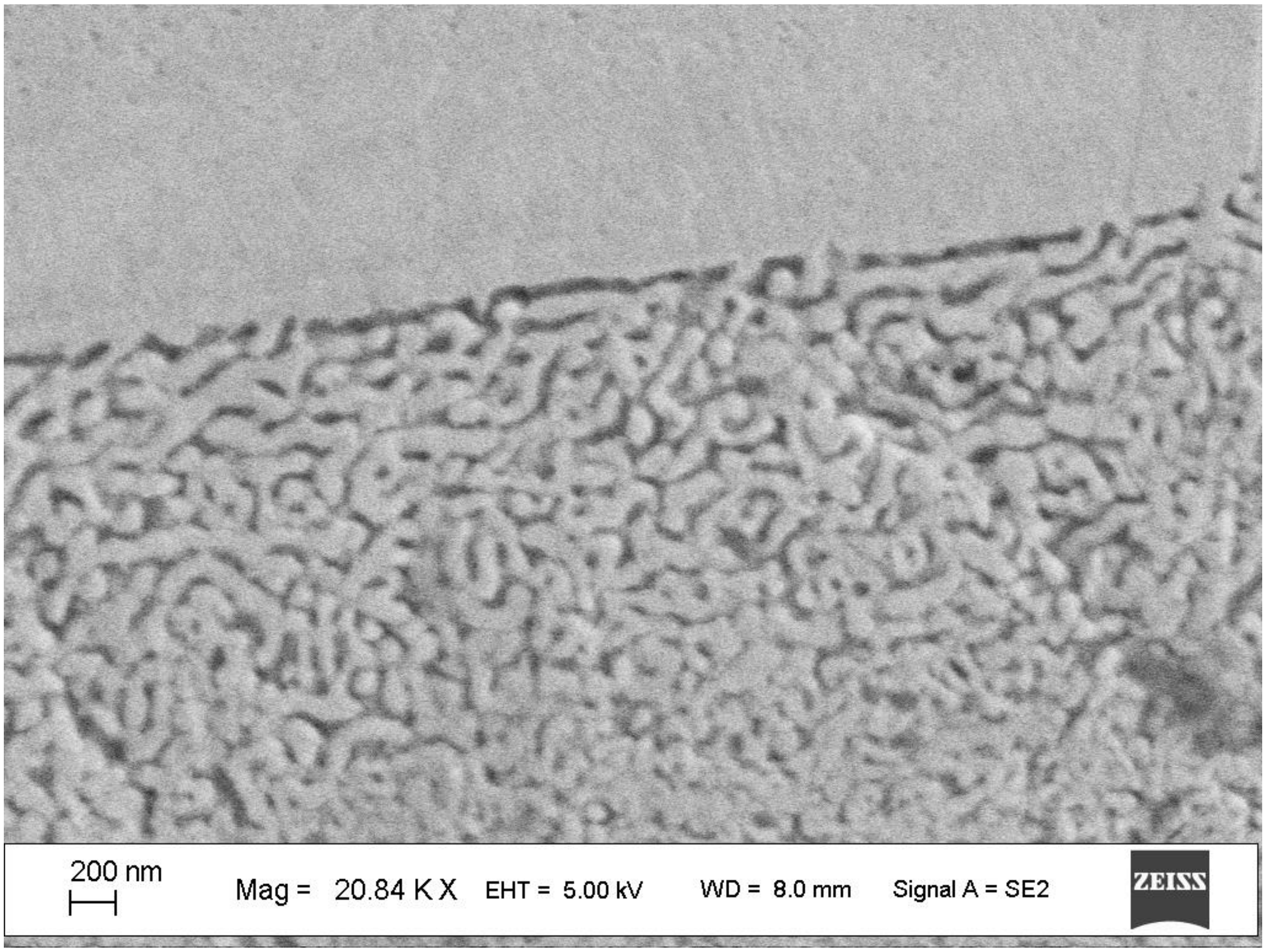

Fig. S1. Scanning electron microscope image of the tetrataenite (upper part) – cloudy zone (lower part) microstructure in zoned metal particle.

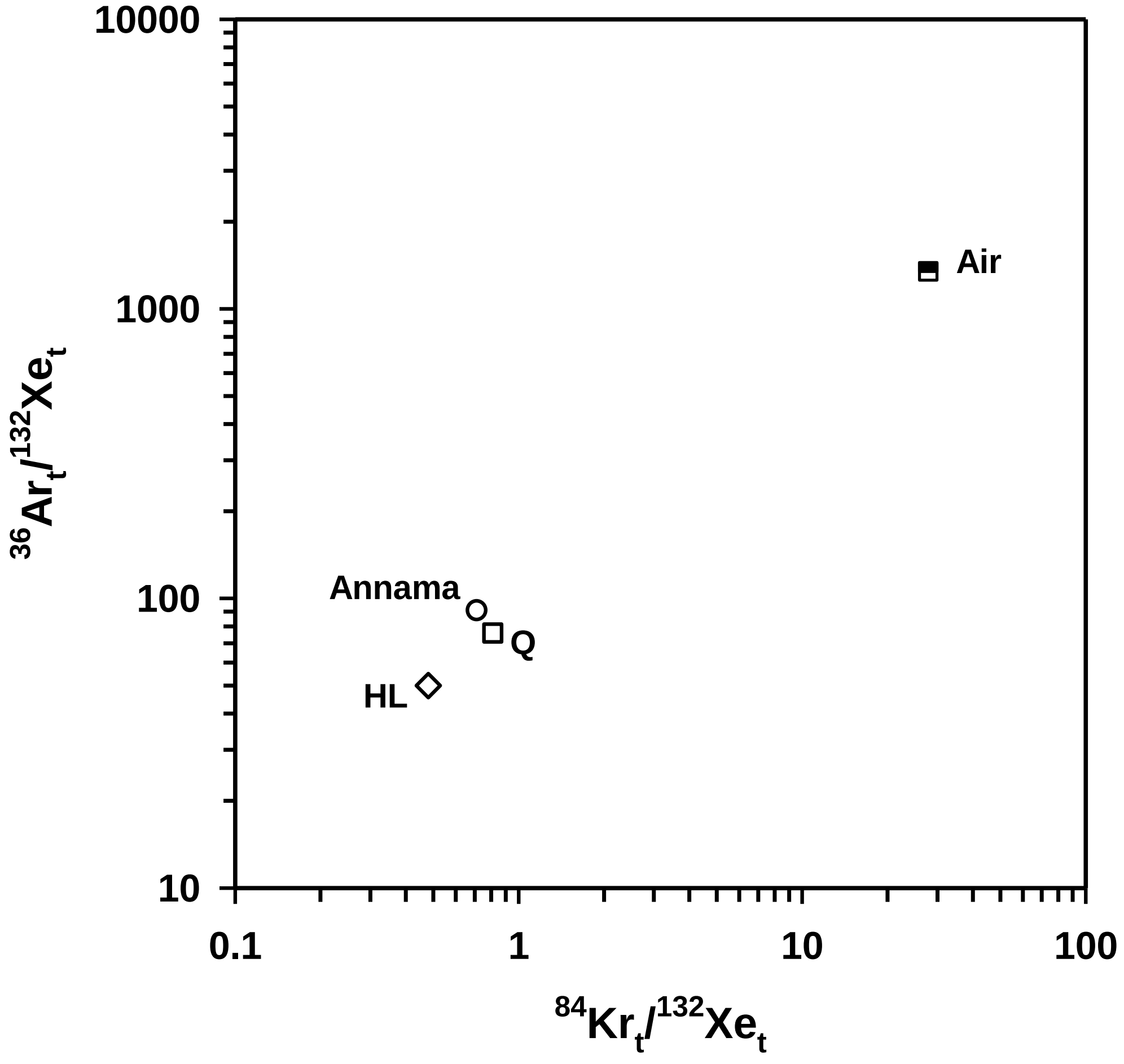

Fig. S2. $^{36}Ar/^{132}Xe$ vs $^{84}Kr/^{132}Xe$ elemental plot of Annama Ir sample. Q, HL and Air components are shown.

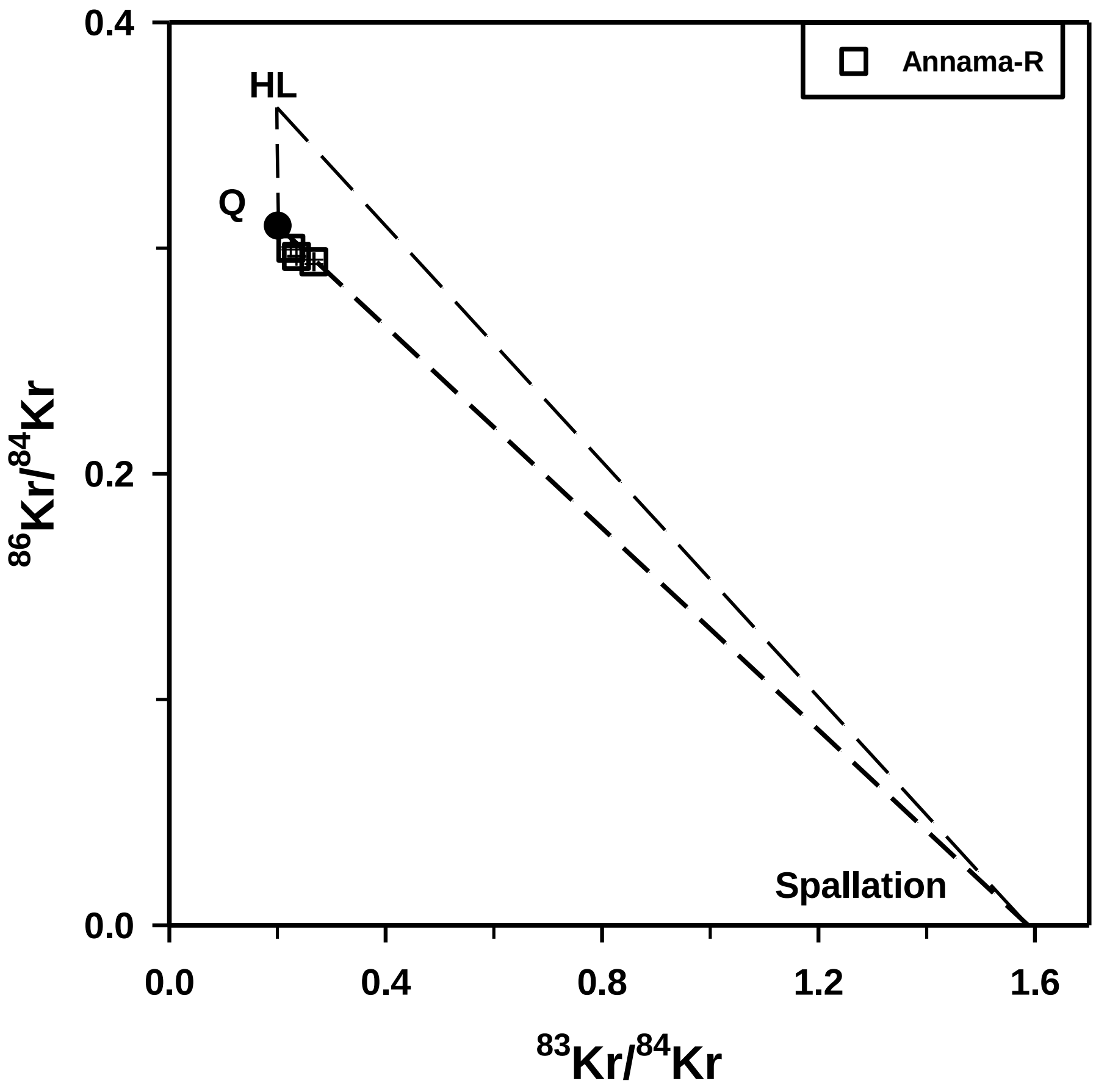

Fig. S3. Krypton three isotope plot of Annama Ir.

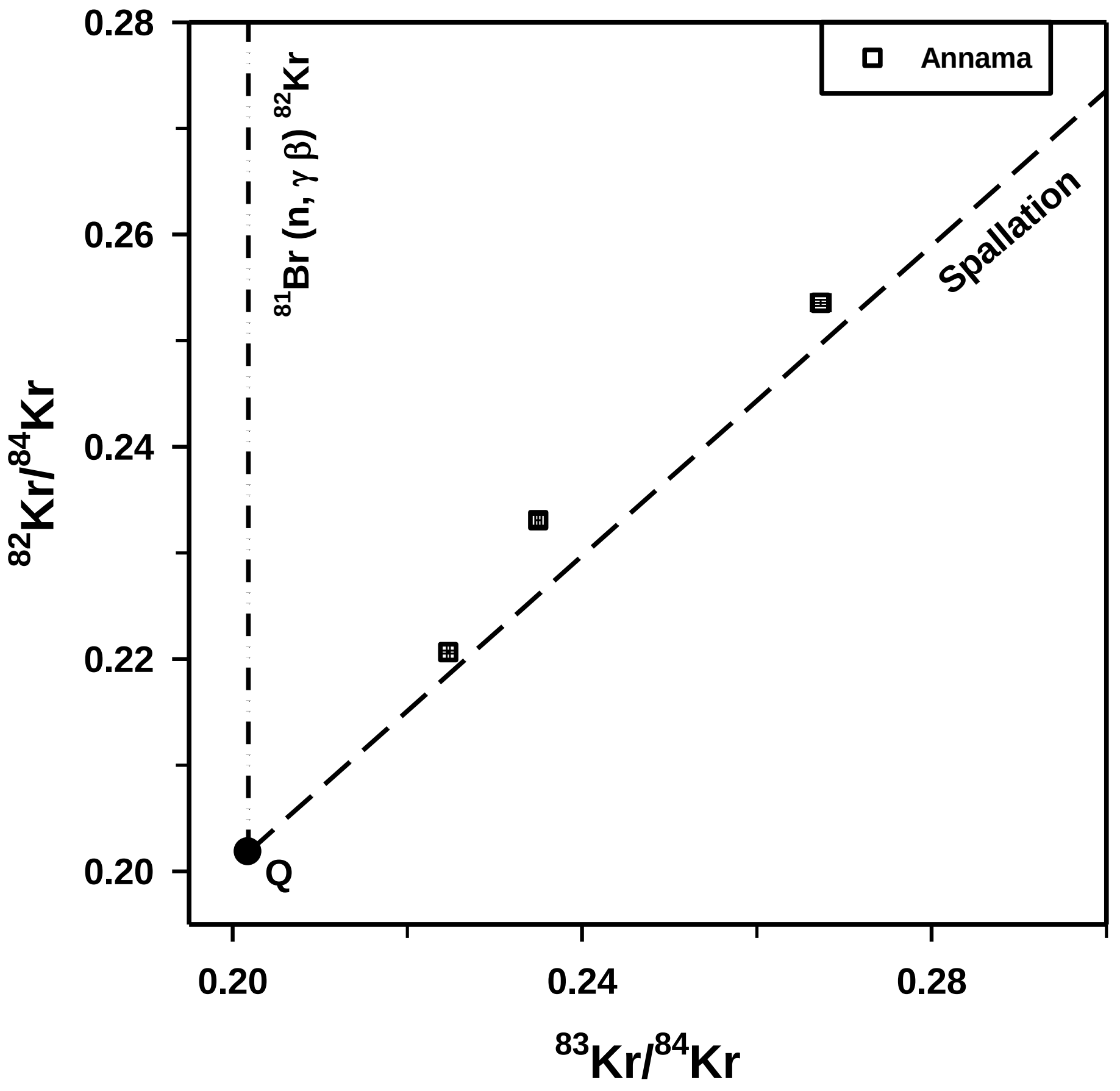

Fig. S4. $^{82}$Kr/$^{84}$Kr vs $^{83}$Kr/$^{84}$Kr three isotope plot of Annama Ir.